\documentclass[aps,prb,twocolumn,superscriptaddress,showpacs]{revtex4-2}
\usepackage{amsmath,amssymb,amsfonts,amsthm}
\usepackage{braket}
\usepackage{graphicx}
\usepackage{bm}
\usepackage{bbm}
\usepackage{color}
\usepackage{booktabs,tabularx}
\usepackage{dcolumn}   
\usepackage{epstopdf}
\usepackage{bbold}
\usepackage[colorlinks=true,linkcolor=blue,urlcolor=blue,citecolor=blue]{hyperref}
\usepackage{commath}
\usepackage[caption=false]{subfig}
\usepackage{amsmath}
\usepackage{soul}

\newcommand{\vs}[1]{\boldsymbol{#1}}
\newcommand{\avg}[1]{\left< #1 \right>} 
\renewcommand{\d}[2]{\frac{d #1}{d #2}} 

\begin{document}

\preprint{APS/123-QED}

\title{Spatiotemporal Quenches in Long-Range Hamiltonians}

\author{Simon Bernier}
\email{simon.bernier@mail.mcgill.ca}
\affiliation{Department of Physics, McGill University, Montr\'eal, Qu\'ebec H3A 2T8, Canada}
\author{Kartiek Agarwal}%
\email{agarwal@physics.mcgill.ca}
\affiliation{Department of Physics, McGill University, Montr\'eal, Qu\'ebec H3A 2T8, Canada}

\date{\today}

\begin{abstract}
Spatiotemporal quenches are efficient at preparing ground states of critical Hamiltonians that have emergent low-energy descriptions with Lorentz invariance~\cite{agarwal2018,mitra2019}. The critical transverse field Ising model with nearest neighbor interactions, for instance, maps to free fermions with a relativistic low energy dispersion. However, spin models realized in artificial quantum simulators based on neutral Rydberg atoms, or trapped ions, generically exhibit long range power-law decay of interactions with $J(r) \sim 1/r^\alpha$ for a wide range of $\alpha$. In this work, we study the fate of spatiotemporal quenches in these models with a fixed velocity $v$ for the propagation of the quench front, using the numerical time-dependent variational principle. For $\alpha \gtrsim 3$, where the critical theory is suggested to have a dynamical critical exponent $z = 1$, our simulations show that optimal cooling is achieved when the front velocity $v$ approaches $c$, the effective speed of excitations in the critical model. The energy density is inhomogeneously distributed in space, with prominent hot regions populated by excitations co-propagating with the quench front, and cold regions populated by counter-propagating excitations. Lowering $\alpha$ largely blurs the boundaries between these regions. For $\alpha < 3$, we find that the Doppler cooling effect disappears, as expected from renormalization group results for the critical model which suggest a dispersion $\omega \sim q^z$ with $z < 1$. Instead, we show that excitations are controlled by two relevant length scales whose ratio is related to that of the front velocity to a threshold velocity that ultimately determines the adiabaticity of the quench. 
\end{abstract}

\maketitle

\section{Introduction}
\label{sec:intro}

Modern quantum simulators hold immense potential for studying fundamental aspects of quantum many-body systems and materials. Recent experiments in ultracold atoms~\cite{bloch2008, bloch2012, gross2017, browaeys2020} and trapped ions~\cite{blatt2012,garttner2017,monroe2021} have successfully demonstrated many novel quantum phenomena---a variety of spin models~\cite{agarwal2020,keesling2019,jepsen2021,scholl2021,ebadi2021,scholl2022}, topological quantum numbers~\cite{tarruell2012,grusdt2013,leseleuc2019,wintersperger2020,semeghini2021}, many-body localization~\cite{pal2010,alet2018,schreiber2015,choi2016,smith2016,morong2021}, lattice gauge theories~\cite{zohar2016,haukeQS2013,martinez2016,banuls2020}, among others~\cite{choi2019,bluvstein2021,viermann2022}---and have emerged as candidates for programmable quantum computing~\cite{kasper2021,ebadi2022}. A key application of such artificial quantum matter is to simulate strongly correlated phases of electrons in conventional materials~\cite{mazurenko2017,linke2018,tarruel2018,bohrdt2021}. Although Hamiltonians of many such systems can be approximately realized using a combination of fixed potentials and driving, it remains a challenge to prepare the system in a state corresponding to a low enough effective temperature at which the ground state properties can be reliably explored~\cite{mazurenko2017}.


Conventionally, state preparation proceeds via adiabatic evolution~\cite{albash2018}. The system is initialized in (or close to) the ground state of a Hamiltonian that is easy to prepare---the Hamiltonian is usually gapped and the ground state has low entanglement. The parameters of the Hamiltonian are then tuned such that the state evolves into the target state, which is often the ground state of a target Hamiltonian. If the parameters are tuned slowly enough, the quantum state stays in the ground state of the instantaneous Hamiltonian. However, the time required to adiabatically prepare a state scales as the square of the inverse of the smallest energy gap encountered when tuning to the target Hamiltonian~\cite{albash2018}. If the gap closes during evolution, excitations are inevitably produced and adiabatic techniques fail to produce the target state with high probability.


In cases where adiabatic evolution fails or takes longer than the coherence time of the quantum simulator, shortcuts to adiabaticity are required. For this purpose, counter-diabatic driving was introduced to counter the production of excitations using auxiliary time-dependent Hamiltonians~\cite{delcampo2012,damski2014,sels2017}. Optimal control protocols such as bang-bang protocols have been developed \cite{pichler2018,ho2019ultra,pagano2020,ebadi2022} and rely on classical optimization of the protocol. Spatially inhomogeneous quenches have also been developed, where portions of a system act as a sink for excitations \cite{ho2009,zaletel2021}.

\begin{figure}
    \centering
    \includegraphics[width=\linewidth]{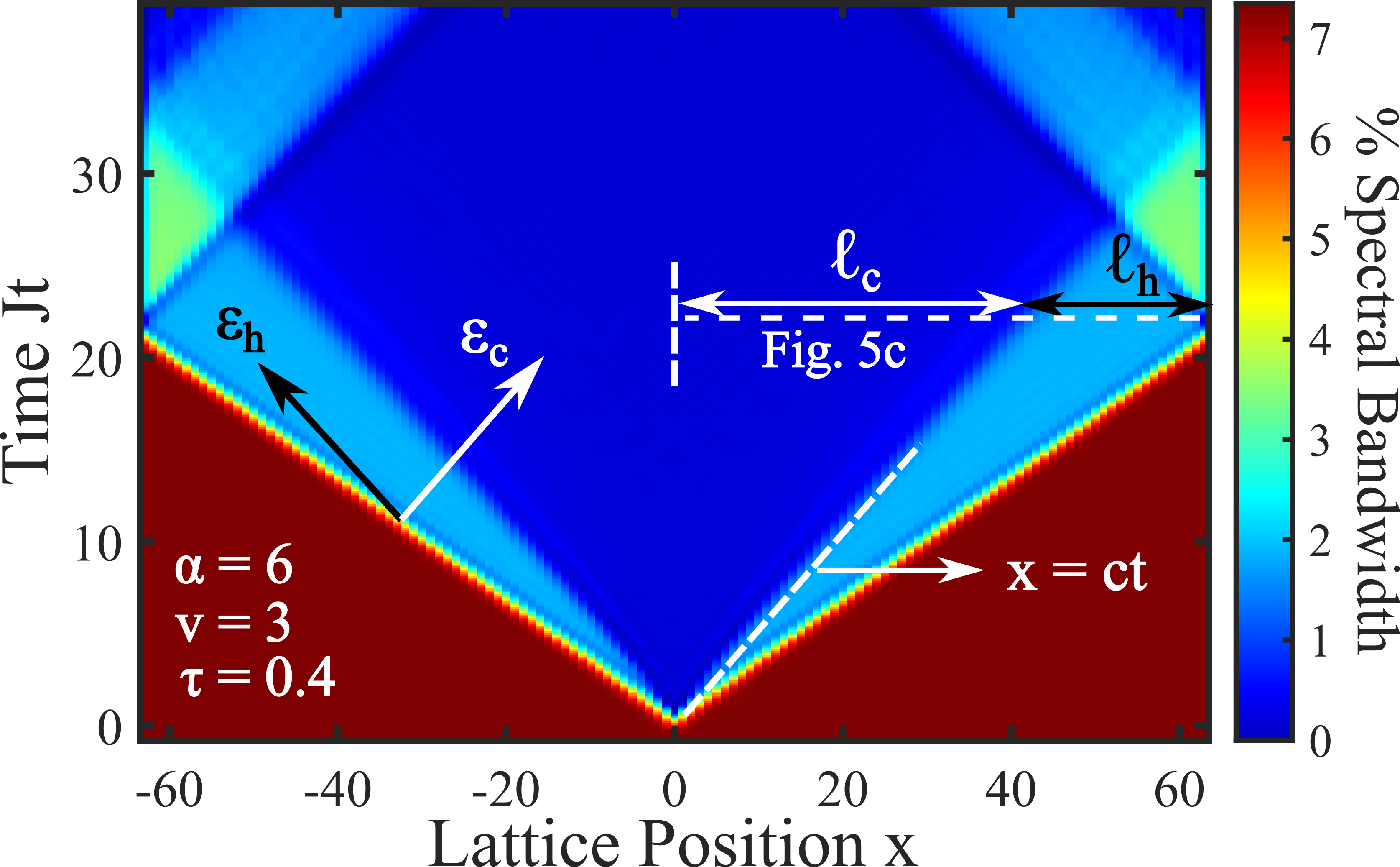}
    \caption{Energy density during a spatiotemporal quench in the LR-TFI model with $\alpha=6$ expressed in terms of the percentage of the critical spectral bandwidth. The quench front acts as a source of excitations, populating modes in a chiral way. At the end of the quench, the modes left in the wake of the front are populated according to a red-shifted temperature leaving a ``cold" region of size $2\ell_c=Lc/v$ at energy $\epsilon_c$. Modes copropagating with the front are populated according to a ``hot" blue-shifted temperature carrying energy $\epsilon_h$ and confined to a region of size $2\ell_h=L(1-c/v)$.}
    \label{fig:colorPlot}
\end{figure}

For systems that exhibit emergent Lorentz symmetry, an efficient route to preparing the ground state of Hamiltonians is via spatiotemporal quenches~\cite{dziarmaga2010,agarwal2017,agarwal2018,mitra2019}. This class of protocols can be used to rapidly produce the ground state of such Hamiltonians even in the critical case, characterized by a linearly dispersing mode with a minimum energy gap that vanishes as $\sim 1/L$, where $L$ is the linear dimension of the system. In particular, the system is initialized in a low-entanglement state corresponding to the ground state of a Hamiltonian that has a gapping perturbation on top of the critical Hamiltonian. The gapping perturbation is then turned off along a quench front moving at a time-dependent velocity $v(t)$ greater than the speed of  ``light" $c$ of the critical theory. In the simplest version of the protocol, $v(t)$ is constant in time and optimal cooling is obtained in the limit $v \rightarrow c^+$; see Fig.~\ref{fig:colorPlot}. Such methods should be applicable to quantum simulators trying to obtain low-energy states of the Hubbard model in two-dimensions~\cite{mazurenko2017,auerbach2012} (in particular at half filling and large $U$, where a linear spin wave dispersion emerges) and in one dimensional quantum gases where a low-energy Luttinger liquid description often applies~\cite{giamarchi2003}. 

Intuitively, the protocol uses Doppler-shifts to result in cooling. In particular, the quench front excites modes in a chiral way. Modes co-propagating with the front are blue shifted while counter-propagating modes are red-shifted. As the velocity of the front approaches the speed of light, counter-propagating excitations are completely suppressed, and all energy is carried by excitations propagating along with the quench front, leaving behind a system with critical ground state correlations. This method thus prepares the ground state of critical models in a time that scales linearly with system size, providing a parametric advantage over adiabatic evolution, which requires a time increasing quadratically with system size~\cite{agarwal2017}.


In this work, we study these spatiotemporal quenches with particular emphasis on an aspect inherent to many artificial simulators based on trapped ion setups or neutral Rydberg atoms. These systems generically realize effective spin models with long-range hopping and interactions, with terms decaying as $1/r^\alpha$ in distance $r$, for $ 0 < \alpha \leq 6$. These long-range interactions (LRIs) inherent to quantum simulators introduce an extra layer of complexity to the effective model realized, and it is vital to understand the effect of LRIs on the efficacy of spatiotemporal quenches to implement them on existing platforms. 

For brevity, we focus our studies on spatiotemporal quenches in one specific model---the long-range transverse field Ising (LR-TFI) model with ferromagnetic power-law interactions and interaction amplitude $J(r) \sim 1/r^{\alpha}$, for a range of $\alpha$. Previous work studying spatiotemporal quenches in the short-range TFI and Heisenberg models~\cite{dziarmaga2010,agarwal2018} whose low energy theories exhibit Lorentz invariance, and thus a maximal speed of propagation of information, and the presence of a linear lightcone that captures causality. We investigate three experimentally motivated values of $\alpha$. We show that for values of $\alpha = \{3,6\}$, where the critical dynamics are relativistic~\cite{maghrebi2016}, `Doppler-shift' cooling persists and approximately the exact results for free fermions if ultraviolet (UV) modes remain unexcited by the quench front. Furthermore, our simulations reveal that there is a clear local minimum in the energy density of excitations as a function of the velocity of the quench front at $v \approx c$. These results highlight the efficacy of such spatiotemporal quenches and underline the fact that optimal diabatic cooling is obtained in the limit where the quench front velocity approaches the emergent speed of light, $v \rightarrow c$.

For $\alpha=2$, the critical dynamics are non-relativistic; the protocol accordingly loses its Doppler cooling effect. Instead, we focus on identifying the relevant length scales that control correlations and excitations---we find that both the usual QKZM length $\xi_{\text{KZ}}$ that governs the formation of defect density in a homogeneous quench, and a length scale $\xi_{\text{SP}}$ that governs relaxation of correlations near a domain wall separating regions on either side of a phase transition, as introduced in Ref.~\cite{dziarmaga2010}, are relevant. This generalized QKZM can be used to predict the behavior of correlations and the energy density of excitations as the velocity of the quench front is varied; we illustrate this using appropriate scaling collapses. These generalized QKZM arguments in principle also apply to the relativistic case but the scaling function itself has non-trivial behavior that can only be predicted using an understanding of Doppler shifts. Importantly, these Doppler shifts predict a local minimum in the excitation of the system in the realitivistic case around $v \approx c$; such a minimum is entirely absent for $\alpha = 2$. 


This manuscript is organized as follows. In Sec.~\ref{sec:models}, we introduce the models studied and their critical properties. In Sec.~\ref{sec:lorentzCooling}, we show that the energy density and correlation length at the end of the quench are qualitatively consistent with Doppler cooling for $\alpha=3$ and $\alpha=6$. We compare our findings to exact calculations computing quench dynamics in a system of free relativistic fermions; these calculations are relegated to App.~\ref{sec:freefermi} for readability of the main text. In Sec.~\ref{sec:alpha2}, the case $\alpha = 2$ is studied, where using appropriate scaling collapses, we identify the length scales governing excitations and correlation functions. We discuss the growth of entanglement entropy during the quench at the end of each section. We conclude with a summary of findings and potential future directions in Sec.~\ref{sec:conclusion}. 

\section{Models studied and Quantum Kibble-Zurek scaling}
\label{sec:models}

We study spatiotemporal quenches in one-dimensional LR-TFI models with Hamiltonians
\begin{equation} \label{eq:Hamiltonian}
    H = -J\left(\sum_{i<j} \frac{\sigma_i^x \sigma_j^x}{|i-j|^\alpha} + g_c \sum_i \sigma_i^z \right) - h\sum_i f_i(t) \sigma_i^z,
\end{equation}
where $\sigma_i^\mu$ are the Pauli matrices, $J$ is the interaction strength, $g_c$ is the critical transverse field and $h$ is the initial gapping perturbation. In what follows, we set $J=1$ and let $h=4$ for all three systems. The perturbation is quenched along smooth fronts moving at velocity $v$ such that $f_i(t) = \frac{1}{2} + \frac{1}{2}\tanh\!\left[(\abs{x_i}-vt)/v\tau\right]$, where $\tau$ is the smoothing parameter. The quench is started at time $t_0=-2\tau$ ensuring that $f_i(t_0)\approx 1$ at every site. At $t = \infty$, $f_i (t) = 0 \forall i$, and the critical Hamiltonian is obtained. The quench time is halved by starting the quench in the center of the chain. In this paper, we restrict the study of quenches in chains of up to $N=256$ spins with all time-dependent results presented for $N=128$.

The system is initialized in the ground state of the paramagnetic phase with large $h$. The initial wavefunction is obtained using ITensor's density matrix renormalization group (DMRG) algorithm~\cite{itensor}. Concretely, the Hamiltonian is represented as a matrix product operator (MPO) consisting of a sum of $k$ exponentially decaying Hamiltonians with different decay lengths that together approximate power-law interactions~\cite{Crosswhite2008,Pirvu2010}. The error in the long-range interaction amplitude at any site 
is restricted to $10^{-6}J$ by using $k= 3, 6, 8$ exponentials for $\alpha = 6$, $3$ and $2$, respectively. The quality of the approximation is verified by a scaling collapse of the energy gap for all three systems---see App.~\ref{sec:properties}---where we find that the critical transverse fields and critical exponents are consistent with previous numerical studies~\cite{zhu2018,koziol2021}. Next, time evolution is carried out with the fourth-order time-dependent variational principle~\cite{haegeman2011,haegeman2016}. At every time step, the spin correlations, the (von Neumann) entanglement entropy and the total energy with respect to the critical Hamiltonian are calculated using standard matrix product state (MPS) techniques~\cite{schollwock2011}.

We also compute the energy density; we define this locally over each bond between sites $i, i+1$, as the expectation value of the operator
\begin{align}
    h_{i} = &-\sum_{\text{odd } r\geq1} \frac{1}{r^\alpha} \sigma^x_{i-\frac{r-1}{2}}\sigma^x_{i+\frac{r+1}{2}} \nonumber \\
    & - \frac{1}{2}\sum_{\text{even } r\geq2} \frac{1}{r^\alpha} \left( \sigma^x_{i-\frac{r}{2}} \sigma^x_{i+\frac{r}{2}} + \sigma^x_{i-\frac{r}{2}+1}\sigma^x_{i+\frac{r}{2}+1} \right)\nonumber \\
    & - \frac{g_c}{2} \left( \sigma^z_i + \sigma^z_{i+1} \right)
\end{align}
where $i\in [1,N-1]$. Here we used the critical value of the transverse field as we are interested in finding the energy above the ground state of the critical system. Note further that $\sum_i h_i = H - \frac{g_c}{2} (\sigma^z_1 + \sigma^z_N$; thus, the local energy density defined in this way sums to the total energy besides irrelevant boundary terms. 

The choice of $\alpha$ studied in this work is motivated experimentally. Dipole-dipole interactions and Van der Waals interactions between neutral Rydberg atoms in optical traps naturally yield effective spin models with $\alpha = 3$ and $\alpha=6$ power-law interactions \cite{browaeys2016}, respectively. Trapped ion experiments can simulate effective spin models with phonon mediated long-range interactions with $0<\alpha<3$. In practice, experiments are limited to $\alpha \lesssim 1.5$~\cite{monroe2021}. We note that in general one-dimensional spin models, the tightest Lieb-Robinson bounds for $1 < \alpha < 2$ predict a logarithmic light cone with a boundary $t \sim \log(r)$, polynomial light cones with $t\sim r^\kappa$ for $2 < \alpha \le 3$~\cite{tran2021} and linear light cones $t\sim r$ for $\alpha>3$~\cite{kuwahara2020}, with $\alpha = 2$ and $\alpha = 3$ being limiting cases of these three regimes. 

The LR-TFI model has a second order phase transition separating a ferromagnetic phase for $g<g_c$ and paramagnetic phase for $g>g_c$. The critical points of the Hamiltonians in Eq.~(\ref{eq:Hamiltonian}) are found by performing a scaling collapse of the energy gap calculated with DMRG as a function of the transverse field, for various system sizes up to $N = 192$; details are presented in App.~\ref{sec:properties}. We find that the calculated critical fields $g_c$ and critical exponents $\nu$ agree well with previous quantum Monte Carlo and DMRG investigations~\cite{zhu2018,koziol2021}. Moreover, renormalization group (RG) calculations of the LR-TFI model with $\alpha=1+\sigma$ predict relativistic dynamics ($z=1$) for $\sigma>7/4$. The critical theory for $\alpha \geq 3$ is understood to be the same as the short range model, thus described by a free fermion theory. For $2/3<\sigma<7/4$, the dynamical critical exponent is calculated in an epsilon series expansion with $\epsilon=3\sigma/2-1$, giving $z = \sigma/2 + \rho(\sigma)\epsilon^2+O(\epsilon^3)$ with $\rho(\sigma)\approx 1/[24(1+\sigma^2)]$. For $\alpha=2$, $z\approx0.505$ and the critical dynamics are non-relativistic.

Near criticality, the equilibrium properties of the system become universal and are described by a correlation length $\xi\sim |g-g_c|^{-\nu}$ where $\nu$ is the critical exponent $\nu$ and $g-g_c$ measures the distance to criticality. The corresponding energy gap is $\Delta \sim |g-g_c|^{z\nu}$, where $z$ is the dynamical critical exponent. In homogeneous phase transitions, where the parameter $g(t)$ is globally tuned to or across the critical value $g_c$ at a rate $1/\tau$, the Kibble-Zurek mechanism describes the typical length- and time-scales at which adiabatic evolution breaks down~\cite{zurek1985,zurek1993,zurek1996}. 
Adiabatic evolution stops when the instantaneous correlation length diverges at a rate faster than a threshold velocity. For a general dispersion relation $\omega \sim q^z$, the group velocity of modes at length scales $\xi$ (or $q \sim 1/\xi$) is given by $v_q \sim \xi^{1-z}$. Solving $d\xi/dt = v_q$ gives $\xi_\text{KZ} \sim \tau^{\nu/(1+z\nu)}$ called the healing length of the system. It describes the typical size of symmetry broken regions caused by the excitations produced during the quench. The threshold velocity can be estimated~\cite{dziarmaga2010} as 
\begin{equation}\label{eq:vstar}
    v^* \sim v_q (q = 1/\xi_{\text{KZ}}) \sim \xi_\text{KZ}^{1-z} \sim \tau^{(1-z)\nu/(1+z\nu)}.
\end{equation}
In systems with $z=1$, the threshold velocity is a constant and given by the maximum group velocity of quasiparticle excitations. The system evolves adiabatically when the correlation length changes at a rate much slower than this threshold velocity (the exact dependence of excitation energy on the quench front velocity may exhibit important non-monotonicity as we discuss in Sec.~\ref{sec:lorentzCooling}).

In inhomogeneous phase transitions, the QKZM must be generalized to account for the moving quench front~\cite{dziarmaga2010}. At fixed time $T$, the quench front is at location $x_c = vT$ and $g(x) \sim g_c + \frac{x-x_c}{v\tau}$ near the front. Even at equilibrium, the correlation length diverges as $\xi(x) \sim g(x)^{-\nu}$, which influences the behaviour of correlations near the front. In a static system with an inhomogeneous perturbation $g(x)$, a length scale $\xi_\text{SP}$ describing the decay of the order parameter to equilibrium in the symmetry broken phase~\cite{zurek2008,damski2009} can be found by comparing the instantaneous correlation length $\xi(x)$ with the distance $x-x_c$ to the critical point on the quench front. This yields
\begin{equation}
    \xi_\text{SP} \sim (v\tau)^{\nu/(1+\nu)}.
\end{equation}


The correlation functions and the energy density in our system generally depend on several length scales---the prominent ones being $\xi_{\text{KZ}}$ and $\xi_{\text{SP}}$---and the spatial region in which the correlator is being calculated. We attempt to identify these length scales in the various cases studied and provide scaling functions where appropriate in the following sections.

\begin{figure}
    \includegraphics[width=\linewidth]{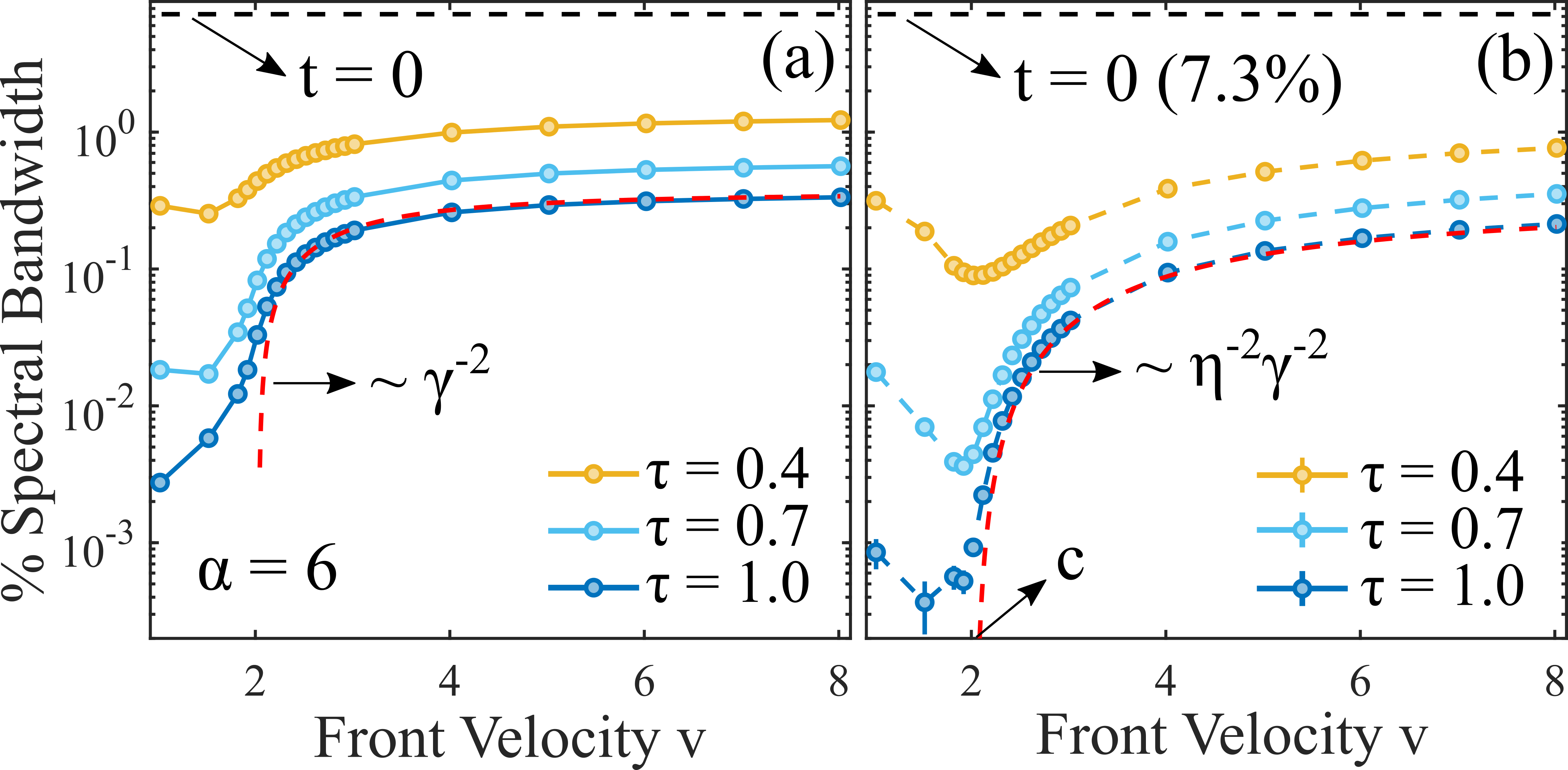}
    \caption{\label{fig:Ealpha6} (a) The total energy density for $\alpha=6$ at time $t_q$ is shown. (b) The average energy density in the center of the spin chain is shown. The averaging is done over the cold region $-R<x<R$, where $R =\min(N/4,ct_q)$ centered about $x=0$ and over a time interval $T = 2$. The red-dotted lines show the free fermion prediction with $c=2.036$ and the black dotted line shows the energy density at the beginning of the quench.}
\end{figure}

\section{Doppler Cooling of Long-Range Models with $z \approx 1$}
\label{sec:lorentzCooling}

The heatwave picture developed in Ref.~\cite{agarwal2017} summarizes the relativistic (or Lorentz) cooling mechanism for superluminal quench fronts presented in this section. In the heatwave picture, the population of modes excited by the quench front is approximated by spatially segregated thermal distributions as shown in Fig.~\ref{fig:colorPlot}. Modes copropagating with the front are confined to the region $ct<\abs{x}<vt$ and populated at a blue-shifted temperature, while the counterpropagating modes occupying the region $0<\abs{x}<ct$ are populated at a red-shifted temperature. For free relativistic fermions, as pertains to the short-range limit of the critical TFI model, one can perform detailed calculation of these mode populations; see App.~\ref{sec:freefermi} for details. The Pauli exclusion principle here prevents the $\sim1/k$ population of modes calculated in the case of spatiotemporal quenches in free bosons~\cite{agarwal2017,agarwal2018}. Instead, the population of modes is given by
\begin{equation}
    N^F_\theta (k) \approx \frac{1}{2} \; \; \text{for} \; \; c k \ll \frac{m}{\gamma \eta (\theta)},
\end{equation}
where $\gamma = 1/\sqrt{1-\beta^2}$ is the Lorentz factor with $\beta=c/v$, $\eta(0) \equiv \eta = \sqrt{\frac{1+\beta}{1-\beta}}$ is the relativistic Doppler factor associated with copropagating modes and $\eta(\pi)=1/\eta$ is associated with counterpropagating modes. In the hot region, modes are occupied up to $k\sim \frac{m}{c}\frac{\eta}{\gamma}$ while those in the cold region are occupied up to $k\sim \frac{m}{c}\frac{1}{\eta\gamma}$. As $v\rightarrow c^+$, the Doppler factor diverges and the cold region is left completely unexcited. Integrating over momenta yields the result that the energy density carried by right and left moving modes is $\sim 1/\left[\gamma^2 \eta (\theta)^2\right]$.
\begin{figure}
    \includegraphics[width=0.95\linewidth]{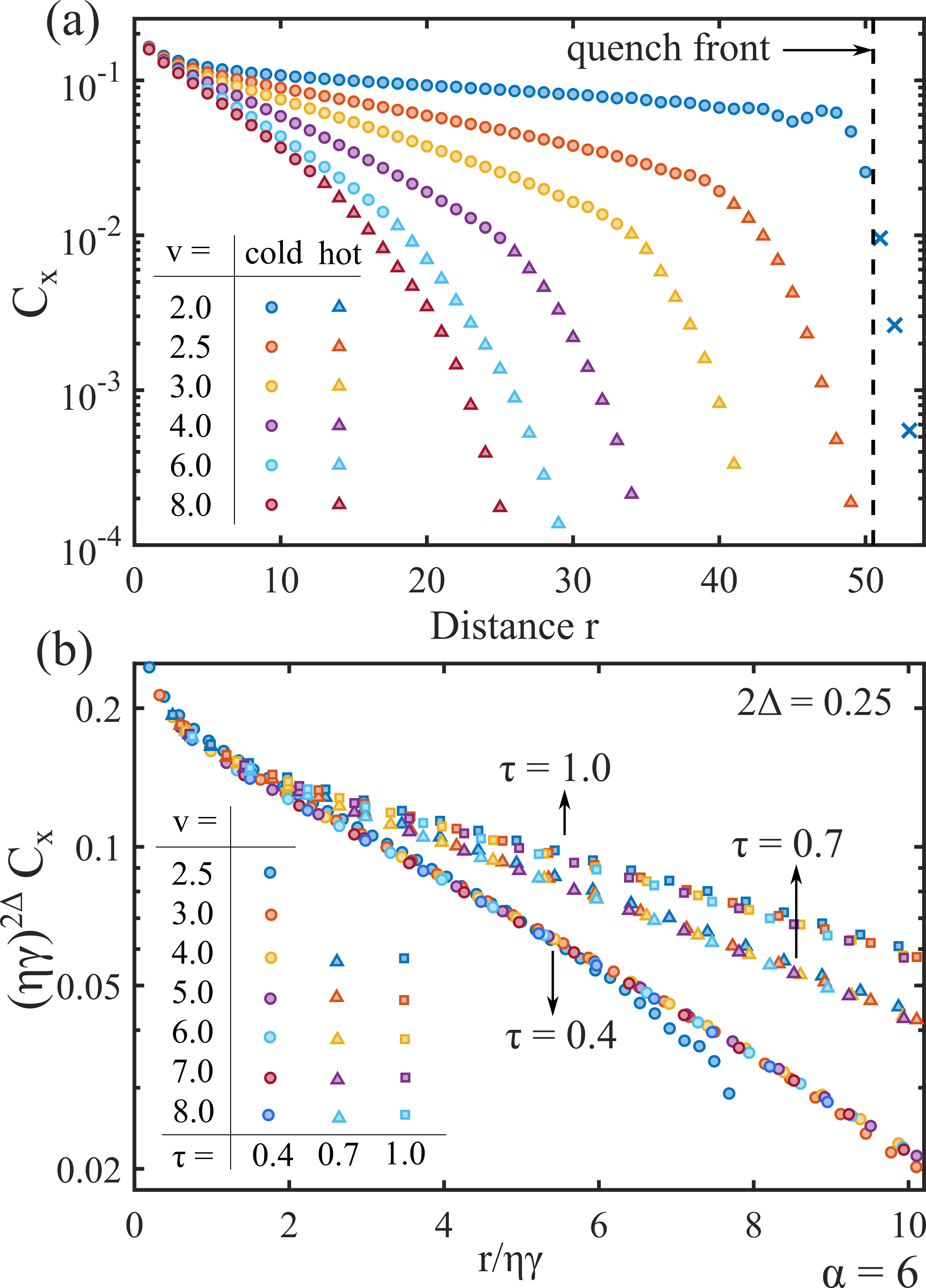}
    \caption{\label{fig:SCalpha6} (a) The spin correlations for $\alpha=6$ are shown for $\tau=0.4$ and various velocities once the quench front reaches $x \approx 50$. The cold region is identified with circles while the hot region is identified with triangles. Crosses seen for $v=2$ represent sites that are still in the gapped region locally (and thus have a decay length different from the cold and hot regions altogether). (b) A scaling collapse of the spin correlations with $\xi = \eta\gamma$ is shown for different $\tau$. The critical exponent of the correlation $2\Delta=0.25$ is found by doing a scaling collapse of the critical correlations (see appendix~\ref{sec:properties}). Note that for $\tau = \{0.7, 1\}$, the velocities $v < 4.0$ are excluded from the scaling collapse as the correlation length approaches $N/2$ and we are limited by finite size effects.}
\end{figure}
We note further that even in the nearest neighbor limit of the TFI model, where an  exact free fermion description holds, the mode dispersion deviates from the relativistic form in the limit $k \rightarrow \pi$. These modes will are not cooled with the above Doppler factors as their group velocity deviates significantly from $c$. Thus, it is necessary to consider a smoothing parameter $\tau$ (which is finite) that prevents UV modes from getting excited---these are expected to be exponentially suppressed for $c k\gg \tau \sim \mathcal{O} (1) $ which prevents heating at energies where the dispersion relation deviates from the linear relation $\omega = c k$.  

We begin by presenting results of spatiotemporal quenches for the LR-TFI models with $\alpha=6$. A scaling collapse of the energy gap reveals that the critical exponents exactly match those of free fermions ($z=1$, $\nu=1$) --- see App.~\ref{sec:properties}. The system is quenched to criticality with $g_c(N=128)=1.01$. Note that $g_c(\infty)=1.031$ reported in App.~\ref{sec:properties} corresponds to the critical value in the thermodynamic limit. 
The excitation energy density at the end of the quench at time $t_q = N/2v + 2\tau$ closely follows the theoretical prediction for free fermions with speed of light $c=2.036$ as shown in Fig.~\ref{fig:Ealpha6}. For free fermions, (which should closely describe the critical properties of the LR-TFI models with $\alpha \gtrsim 3$) we can compute exactly the energy and spatial distribution of excitations due to the spatiotemporal quench. In particular, the energy density $\epsilon_c$ in the cold region $\abs{x}<ct_q$ is found to be $\epsilon_c \sim 1/\eta^2\gamma^2$ while that in the hot region $ct_q<\abs{x}<vt_q$ is $\epsilon_h \sim \eta^2/\gamma^2$ --- see App.~\ref{sec:freefermi}. The average energy density over the entire length of the system must then be $\epsilon_{\text{avg}} \approx \left(1-\frac{c}{v}\right) \epsilon_h + \frac{c}{v} \epsilon_c \sim \frac{1}{\gamma^2}$, as shown in Fig.~\ref{fig:Ealpha6}(a). Near $v=c^+$, the quench excites higher frequency modes that are less efficiently cooled because of their nonlinear dispersion. Increasing $\tau$ restores the cooling effect by adiabatically suppressing the excitation of modes $c k\gg\tau^{-1}$ in the laboratory frame. This effect is particularly strong in the cold region, where the energy density follows $\sim1/\eta^2\gamma^2$ more closely as $\tau$ is increased as shown in Fig.~\ref{fig:Ealpha6}(b).

\begin{figure}
    \includegraphics[width=\linewidth]{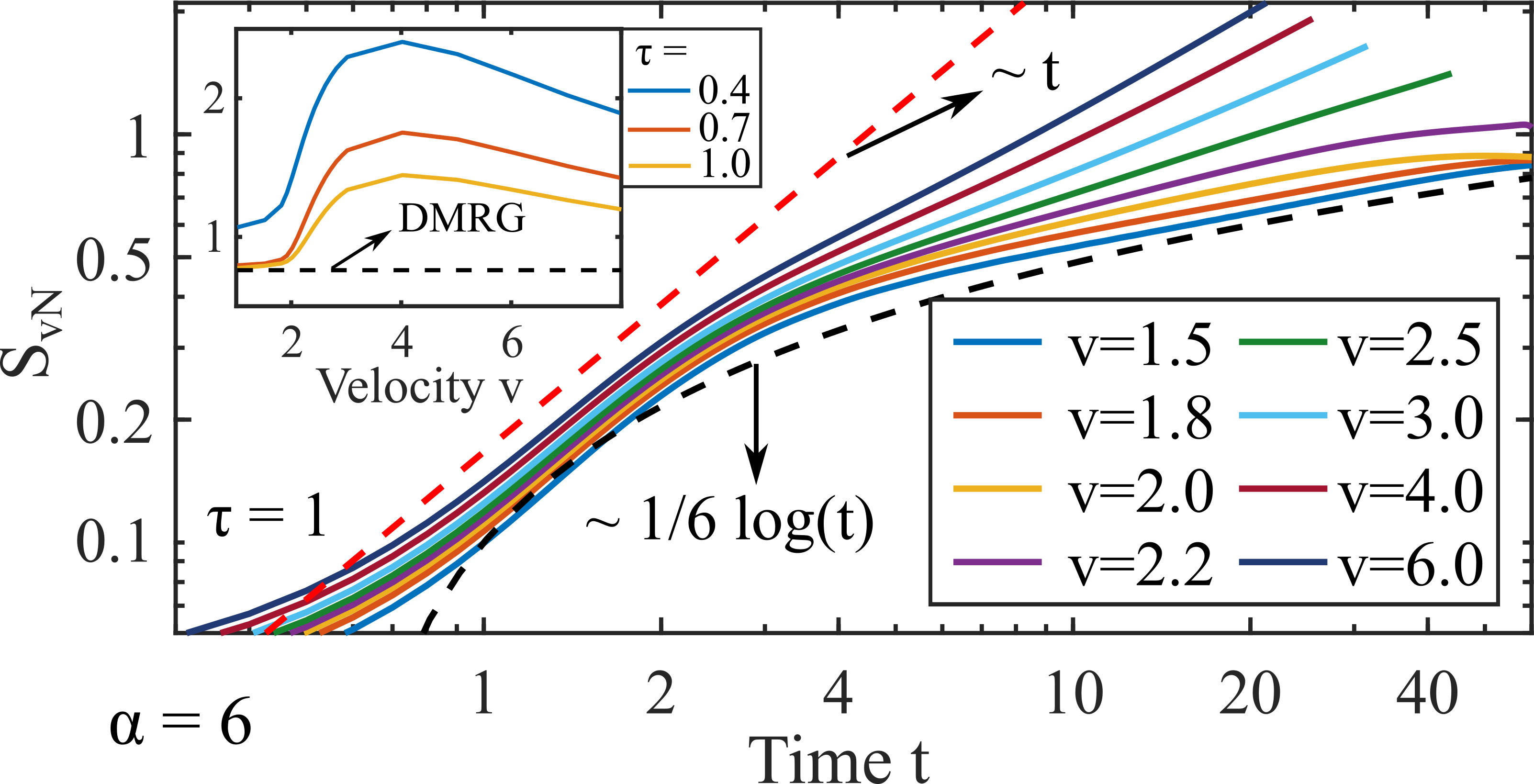}
    \caption{\label{fig:SvNalpha6} The growth of the von Neumann entanglement entropy during the quench for $\alpha=6$ is shown for $\tau=1$. The red-dotted line shows linear increase and the black-dotted line shows a logarithmic $\frac{1}{6}\log(c t)$ growth, as would be expected for the short range TFIM in its ground state over a region of size $L = ct$. Inset: the entanglement entropy at the end of the quench is shown for different $\tau$ as a function of front velocity. The black-dotted line shows the entanglement calculated at criticality using DMRG.}
\end{figure}

\begin{figure}
    \centering
    \includegraphics[width=0.85\linewidth]{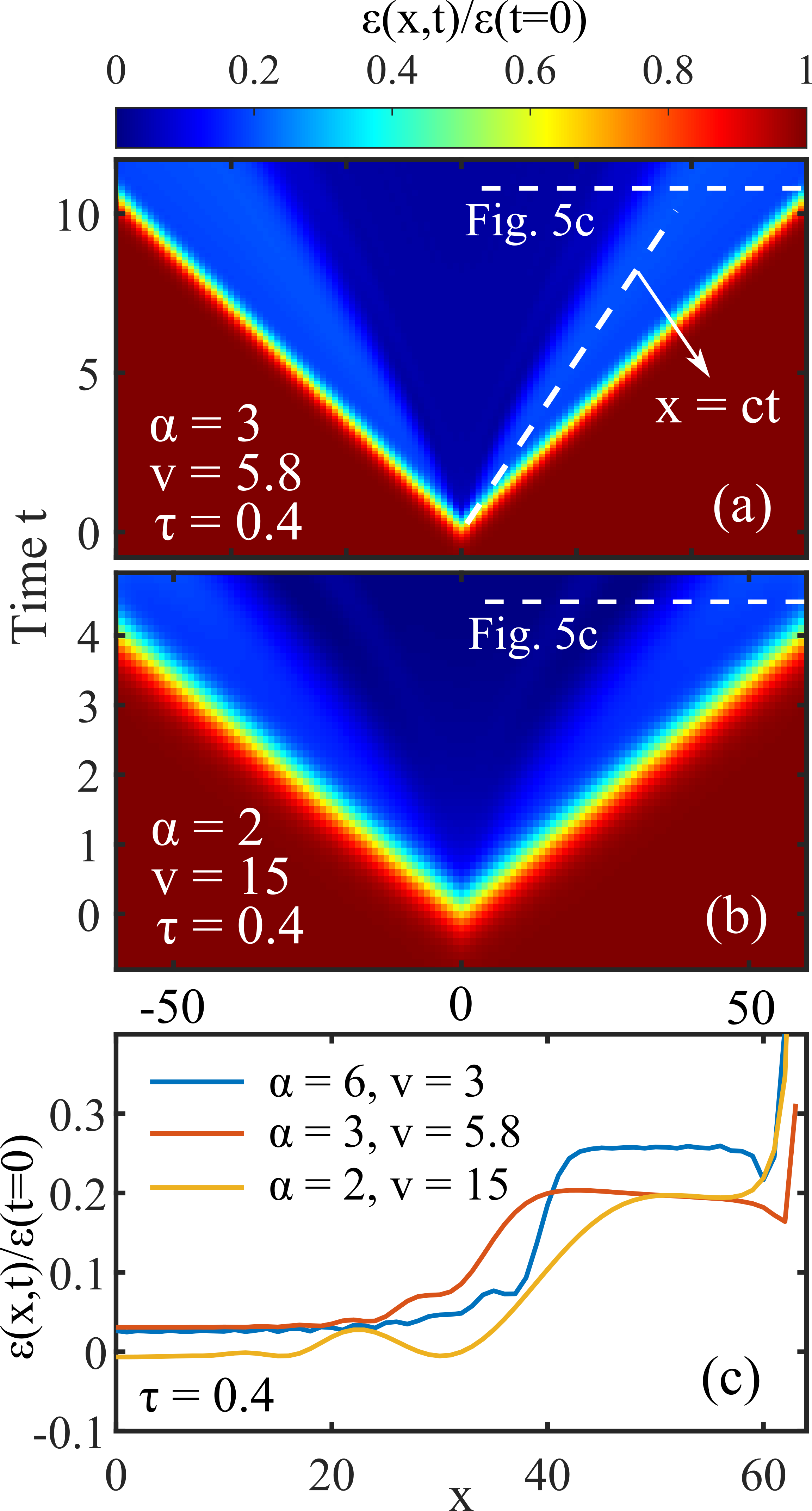}
    \caption{The energy density during a spatiotemporal quench shown as the ratio of $e(x,t)/e(t=0)$. (a) Quenches in systems with $\alpha=3$ show a persisting heatwave picture. The white dotted line shows the light cone with $c$ estimated via a collapse of the spin correlations. The initial energy is $e(t=0)=4.9\%$ of spectral bandwidth. (b) For $\alpha=2$, large front velocities are necessary to see a separation of higher and lower energy density regions. This is due to the existence of the dynamical threshold velocity $v^*$ and is not a relativistic effect. Initially, $e(t=0)=1.8\%$ of spectral bandwidth. (c) Comparing $e(x,t_q)/e(t=0)$ for different $\alpha$.}
    \label{fig:heatwave}
\end{figure}

\begin{figure*}
    \includegraphics[width=\linewidth]{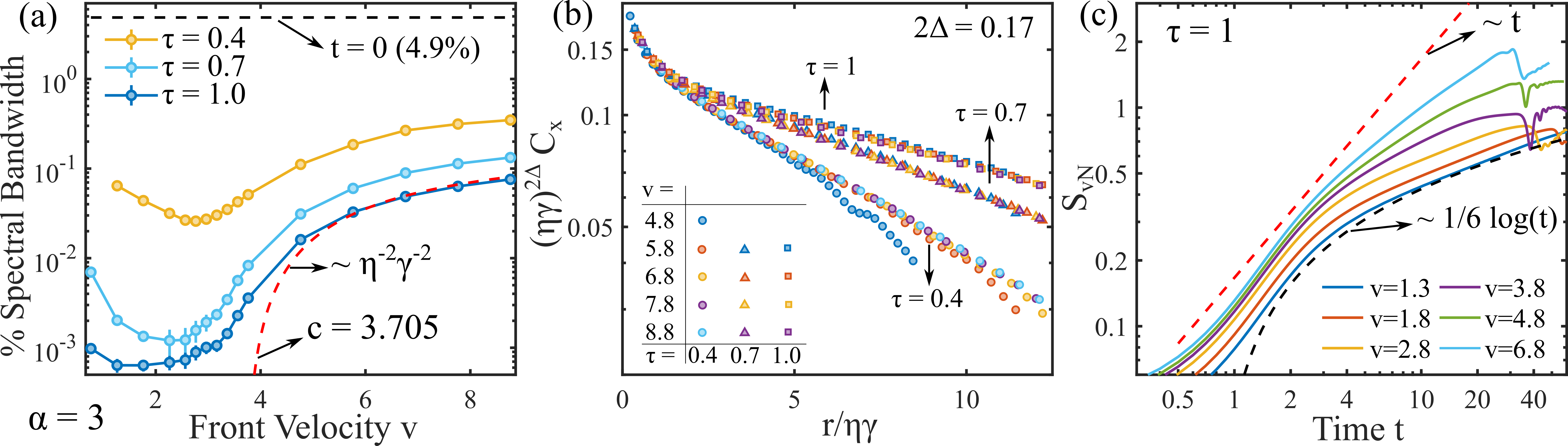}
    \caption{\label{fig:alpha3} (a) The average energy density for $\alpha=3$ at time $t_q$ is shown. The averaging is done over the cold region $-R<x<R$, where $R =\min(N/4,ct_q)$ centered about $x=0$ and over a time interval $T = 2$. The red-dotted line shows the free fermion prediction for $c=3.705$ (as found in appendix~\ref{sec:properties} and the black-dotted line shows that initial energy density. (b) The spin correlations are collapsed using $\xi=\eta\gamma$. The best collapse for $c=3.705$ is obtained for $2\Delta = 0.17$, a value smaller than the predicted $2\Delta=0.25$ (see appendix~\ref{sec:properties}). (c) The growth of von Neumann entanglement entropy is shown. The red-dotted line shows linear increase while the black-dotted shows a $1/6\log(t)$ increase.}
\end{figure*}
The apparent minimum that can be seen near $v=c$ in Fig.~\ref{fig:Ealpha6}(b) is due to Doppler cooling in both limits of $v\rightarrow c$. For subluminal quenches, the modes excited by the moving front continue interacting with it during the entirety of the quench. The waves simply bounce back and forth between the right and left propagating quench fronts.
A classical solution to a moving front is presented in App.~\ref{sec:sublumCooling}. It shows that in the limit of $v\rightarrow c^-$, modes reflecting off of the moving front are red-shifted by $1/\eta^2$ while no modes are transmitted. This leads to a $\sim 1/\eta^2$ cooling. Note that the Doppler cooling factor $\eta$ as defined in the subluminal case is the same as in the superluminal case but with $\beta\rightarrow1/\beta$. Both lead to near-perfect red-shifts in the limit $v \rightarrow c$.)

The heatwave picture motivated in Ref.~\cite{agarwal2017} is also supported by the spin correlations. The correlations decay exponentially with two length scales that can be associated with the cold and hot regions as shown in Fig.~\ref{fig:SCalpha6}(a). Using the observation that the excitation energy scales as $\epsilon \sim \xi^{d+z}$ near criticality, the correlation length $\xi_c=\eta\gamma$ can be identified in the cold region (corresponding to an energy density $\epsilon_c$).

The correlator decays away from $x = 0$ on two different length scales corresponding to the cold and hot regions, as seen in Fig.~\ref{fig:SCalpha6}(a). Additionally, the correlation length increases as the quench front velocity $v$ approaches $c$ in the cold region, while it decreases in the same limit in the hot region, which agrees again with the heatwave picture. Quantitatively, we examine the autocorrelator in the cold region using the following ansatz
\begin{equation}\label{eq:equiSC}
    C_x(r<ct) = \langle \sigma_0^x \sigma_{r}^x \rangle = \xi_c^{-2\Delta} F_C\left(\frac{r}{\xi_c}\right),
\end{equation}
which identifies the scaling of the correlation length $\xi_c$ with the quench velocity according to the heatwave picture. We note that the autocorrelation function shows good scaling collapse with $\xi_c$ over multiple quench front velocities, as seen in Fig.~\ref{fig:SCalpha6}(b).

In the spatiotemporal quench we study, the entanglement entropy also appears to increase slower than the expected linear growth in homogeneous quenches~\cite{calabrese2005,calabrese2009} as shown in Fig.~\ref{fig:SvNalpha6}. Optimal quench protocols with $v\rightarrow c$ and large $\tau$ show nearly logarithmic growth of the entanglement and $S_\text{vN}$ at $t_q$ approaches the ground state value calculated at criticality with DMRG, as shown in the inset of Fig.~\ref{fig:SvNalpha6}.

The heatwave picture persists in spatiotemporal quenches in the LR-TFI model with $\alpha=3$, as shown in Fig.~\ref{fig:heatwave}(a). The system is again initialized with $h=4$ and quenched to criticality with $g_c(N=128)= 1.38$. For this system, RG calculations predict that $z=1$ and that critical dynamics are relativistic. However, a fit of the minimum energy gap as a function of the system size gives $z\approx0.9$. The critical field and critical exponents obtained from a collapse of the gap energy using $z=0.9$ closely matches the results reported in Refs.~\cite{zhu2018,koziol2021}. Despite this deviation from $z=1$, the relativistic cooling effect persists as can be seen from the energy density in the center of the chain at $t=t_q$ as shown in Fig.~\ref{fig:alpha3}(a) where a clear minimum is realized as a function of the quench front velocity, and the appearance of hot and cold regions, as seen in Fig.~\ref{fig:heatwave}(c). We note that in principle a $\tau-$dependent threshold velocity $v^*$ as estimated by QKZM arguments [Eq.~(\ref{eq:vstar})] can replace the critical velocity $c$ in a putative heatwave picture that does not rely on relativistic Doppler cooling. However, the velocity $v^*$ as understood in these arguments merely distinguishes quenches which proceed adiabatically (for $v < v^*$) vs. those that proceed non-adiabatically (for $v > v^*$). It cannot explain the \emph{minimum} observed in the energy density as a function of the quench front velocity. The latter requires a Doppler cooling interpretation and suggests the applicability of such relativistic physics even in the $\alpha = 3$ case. 

We note further that the results are not in as close agreement with the free fermion findings as for $\alpha = 6$. The energy density in the cold region appears to follow the free fermion result for velocities $v \gtrsim 5$ with the critical velocity $c = 3.705$ (calculated by fitting a light cone for correlations arising from a local quench; see App.~\ref{sec:properties}) but fails to show good agreement at smaller quench front velocities. Unlike the $\alpha = 6$ case, the interface between the hot and cold regions does not appear to coincide with $\abs{x}=ct$, as seen in Fig.~\ref{fig:heatwave}(a),  with the velocity $c$ as identified above. This is putatively due to the longer range of interactions which also smooth out the separation between hot and cold regions as $\alpha$ is lowered, as seen in Fig.~\ref{fig:heatwave}(c), and make identification of a clear boundary between these regions challenging. A scaling collapse with the correlation length $\xi = \eta\gamma$ is provided in Fig.~\ref{fig:alpha3}(b). The best collapse using $c=3.705$ reveals that $2\Delta=0.17$, which deviates slighlty from the ground state value of $1/4$ calculated from a scaling collapse of the spin correlations (see App.~\ref{sec:properties}). The discrepancy can arise from two sources---we are not precisely at the ground state after these quenches, and the correlation length can exceed the finite system size studied for a range of velocities. 


These results show in summary that the heatwave picture extends even to $\alpha=3$, and the physics of Doppler cooling can be seen even though we obtain a dynamical critical exponent that deviates slightly from $z=1$. 
The entanglement entropy produced during the quench grows slower than linearly as shown in Fig.~\ref{fig:alpha3}(c) and grows nearly logarithmically for $v \lesssim 3$. Together, these data show that the Doppler cooling effect in spatiotemporal quenches continues to offer a strong advantage over homogeneous quenches in preparing critical states of the LR-TFI model for $\alpha\gtrsim 3$ where $z\approx1$.



\begin{figure*}
    \includegraphics[width=\linewidth]{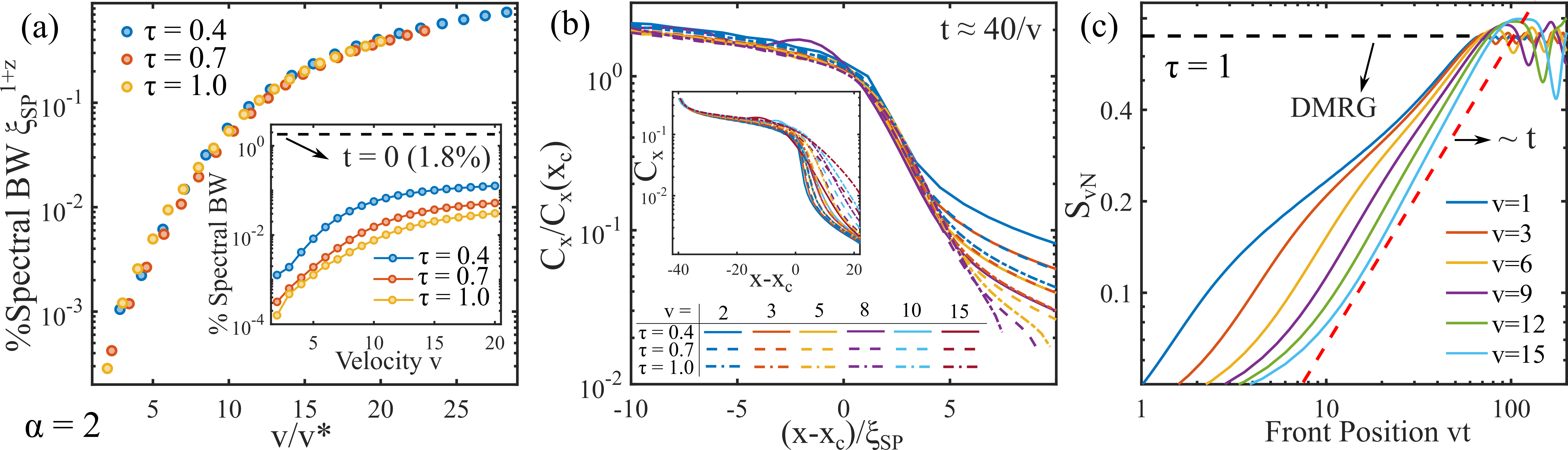}
    \caption{\label{fig:alpha2} (a) The total energy density for $\alpha=2$ at time $t_q$ is shown to collapse according to Eq.~\ref{eq:kzmEnergy}. \textbf{Inset} The energy density at time $t_q$ is shown, with the black-dotted line indicating the initial energy density. (b) The spin correlations near the quench front are collapsed for $v\leq8$, which shows that $\xi_\text{SP}$ is the dominant length scale. The quench front is located at $x_c \approx 40$ in all cases. \textbf{Inset} The uncollapsed correlations are shown, including $v=10$ and $15$. (c) The growth of von Neumann entanglement entropy is shown. The red-dotted line shows linear increase and the black-dotted line shows the entanglement entropy calculated with DMRG at criticality.}
\end{figure*}

\section{Kibble-Zurek mechanism for $\alpha = 2$}
\label{sec:alpha2}


Increasing the range further to $\alpha = 2$ leads to different dynamics. The system is quenched to criticality with $g_c(N=128)=2.3285$. Higher and lower energy regions, as shown in Fig.~\ref{fig:heatwave}(b), are still visible in the quenches although higher velocities are needed to observe this distinction. The presence of hot and cold regions may be explained by the existence of the dynamical threshold velocity $v^*$ in Eq.~(\ref{eq:vstar}). As was the case with $\alpha=3$, there is no sharp interface between these regions as shown in Fig.~\ref{fig:heatwave}(c).

To analyze the properties of the system during the quench, we use the dynamical critical exponent $z\approx0.505$ calculated using RG in Ref.~\cite{maghrebi2016}. The critical theory is suggested to have dispersion modes with a dispersion $\omega \sim q^{1/2}$, and Doppler-shift cooling arguments cannot be applied here. Instead, we show that the dynamics are dominated by a competition between the homogeneous-quench QKZM healing length $\xi_\text{KZ} \sim \tau^{\nu/(1+z\nu)} $, and the static inhomogeneous length scale $\xi_\text{SP}\sim (v\tau)^{\nu/(1+\nu)}$. The ratio of $\xi_\text{KZ}$ and $\xi_\text{SP}$ in fact relates to the ratio of the quench front velocity to the $\tau-$dependent threshold velocity of the quench, 
\begin{equation}
    \frac{\xi_\text{SP}}{\xi_\text{KZ}} \sim \left(\frac{v}{v^*}\right)^{\nu/(1+\nu)}, 
\end{equation}
where $v^* = \tau^{(1-z)\nu/(1+z\nu)}$ is the dynamical threshold velocity identified in Eq.~(\ref{eq:vstar}). 

At fixed time $t^*$, a scaling ansatz for the spin correlations and excitation energy density involving these two length scales yields
\begin{align}
    &\epsilon(t^*) = \xi_\text{SP}^{-(d+z)} F_e\left(\frac{v}{v^*}\right), \label{eq:kzmEnergy} \\
    &\frac{C_x(x,t^*)}{C_x(x_c,t^*)} = F_C\left(\frac{x-x_c}{\xi_\text{SP}}, \frac{v}{v^*} \right), \label{eq:kzmSC}
\end{align}
where $F_e$ and $F_C$ are unknown functions, and the ratio of the length scales $\xi_{\text{KZ}}$ and $\xi_{\text{SP}}$ is considered inside the scaling functions via the ratio $v/v^*$. Note that the scaling ansatz of Eqs.~(\ref{eq:kzmEnergy},\ref{eq:kzmSC}) also works for $\alpha=\{3,6\}$ but only an exact calculation can reveal the Doppler cooling effect that leads to non-monotonic cooling as a function of $v$. 

We begin by performing a scaling collapse of the total energy density at the end of the quench according to Eq.~(\ref{eq:kzmEnergy}) as shown in Fig.~\ref{fig:alpha2}(a). The collapse confirms that the dynamics are influenced by a non unique threshold velocity $v^*$ determined by the quench rate $\tau$. We also note that there does not appear to be any minimum in the energy density as a function of the quench front velocity. The energy density simply decreases as this velocity is lowered, in a marked difference from the result for $\alpha = 3,6$. This further confirms that the $\alpha = 2$ system is not characterized by Doppler cooling. A collapse of the spin correlations is done when the quench front reaches $x=40$ (at time $t=40/v$) according to Eq.~(\ref{eq:kzmSC}), as shown in Fig.~\ref{fig:alpha2}(b). The collapse works reasonably well for $-10 \xi_{\text{SP}} < x-x_c < 5\xi_{\text{SP}}$ for velocities $v<8$ and it shows that $\xi_\text{SP}$ is indeed the correct length scale determining correlations close to the quench front. This length scale also appears to play a role in determining the energy density over the entire spin chain. For $v>8$, $\xi_\text{SP}$ is large enough at $x=40$ that the collapse of correlations does not work well due to strong interaction with the system boundaries, as can be seen in the inset of Fig.~\ref{fig:alpha2}(b).

The von Neumann entanglement entropy also grows differently for $\alpha = 2$ than for $\alpha = 3,6$. The growth is approximately linear in time until the quench front reaches the edges of the spin chain, where it saturates close to the expected value of the entanglement entropy in the ground state of the system (as found from DMRG), as can be seen in Fig.~\ref{fig:alpha2}(c). However, the entanglement entropy appears to show larger oscillations about this mean value for larger velocities of the quench front, in agreement with the reduction in energy density of excitations as the quench front velocity is lowered. [We note in general that the entanglement entropy can be tricky to interpret as low entanglement (as in a product state) and high entanglement (as found for excited eigenstates) both imply heating.] The linear increase of $S_\text{vN}$ is putatively due to the nearly logarithmic light cone $t\sim \log r$ (see Ref.~\cite{tran2021}) quickly spreading information over the entire spin chain. 
It is an open question whether a faster protocol with a time-dependent quench front velocity could be used to create low-energy states in this system. 

The above results largely imply that for $\alpha = 2$, the front velocity serves as another adiabatic parameter with optimal cooling achieved in the limit $v\rightarrow0$; a threshold velocity $v^*$ here can be used as a guideline approximately demarcating adiabatic and non-adiabatic quenches. 


\section{Conclusion}
\label{sec:conclusion}

In this paper, we show that smooth spatiotemporal quenches can efficiently prepare critical ground states of one dimensional LR-TFI models. We confirm that for models with $z\approx1$, when interactions $J(r) \sim 1/r^\alpha$ decay faster than the case $\alpha = 3$, a smooth quench front moving along $x=vt$ leaves a large section of the spin chain unexcited when $v \rightarrow c$, the velocity of excitations in the critical system. This is evidenced by the close resemblance between the energy density computed numerically for these LR-TFI models and that calculated exactly for free relativistic fermions. In particular, the excitation energy shows a local but pronounced minimum as a function of the quench front velocity $v$ for $v \approx c$. For general $v>c$, a heatwave picture emerges where one obtains spatially separated hot and cold regions in the system populated by excitations emanating from the quench front and either co- or counter-propagating with respect to the moving quench front. Going from $\alpha = 6$ to $\alpha = 3$ smooths the separation between these hot and cool regions without qualitatively impacting the results. Simple scaling relations accompanied by a collapse of the spin correlations show that the correlations decay on the diverging length scale $ \xi_c\sim\eta\gamma$. The optimal quench protocol also shows a nearly logarithmic growth of the von Neumann entanglement entropy.

When the dynamical exponent $z$ deviates sufficiently from unity as for $\alpha=2$ [$J(r) \sim 1/r^2$], one still obtains an approximate heatwave picture of excitations in the system for large velocities of the quench front, with areas of high and low energy density; a threshold velocity $v^*$ determined using QKZM arguments potentially serves the role played by the critical velocity in the $z \approx 1$ case in separating hot and cold regions. However, there is no local minimum in  the energy density of excitations at a function of the front velocity, which suggests an important departure from the Doppler cooling picture prevalent in the above cases. Instead, we considered a general scaling picture in this case---we find that energy density and spin correlations are determined by the interplay between two length scales, one corresponding to the size of broken symmetry clusters in a homogeneous quench according to usual QKZM expectations, and another corresponding to the healing length of correlations in a system with a spatially inhomogeneous gap with $g < g_c$ on one side of the system and $g > g_c$ on the other side. The ratio of these lengths is in fact related to the ratio of the velocity of the quench front to a threshold velocity $v^*$ that controls adiabaticity of the quench. The state prepared grows monotinically closer to the target critical ground state as the velocity of the quench front is reduced. 


This paper motivates the implementation of spatiotemporal quenches in one-dimensional spin chains on modern quantum simulators. It remains to be shown that Doppler cooling persists for two-dimensional systems with non-linear dispersions. Spatiotemporal quenches are not expected to provide a cooling advantage in all long-range systems. For power-law interactions with $\alpha\leq1$, Lieb-Robinson bounds do not exist and as such, there is no light cone~\cite{tran2021}. In that case, we could expect that the optimal protocol is homogeneous as argued in Ref.~\cite{ho2019ultra} concerning the preparation of the critical state of the fully connected ($\alpha=0$) TFI model. It however remains to be understood whether a quench with a time-dependent front velocity $v(t)$ (see, for instance, Ref.~\cite{mitra2019}) could be used to optimize the speed and efficacy of ground state preparation for models with $z < 1$ where some kind of light cone exists.

\begin{acknowledgments}
The authors acknowledge useful discussions with several previous collaborators on related work. SB acknowledges support from an FRQNT graduate scholarship. KA acknowledges support from the NSERC Discovery Grant and an INTRIQ team grant from the FRQNT. 
\end{acknowledgments}

\appendix

\begin{figure*}
    \includegraphics[width=\linewidth]{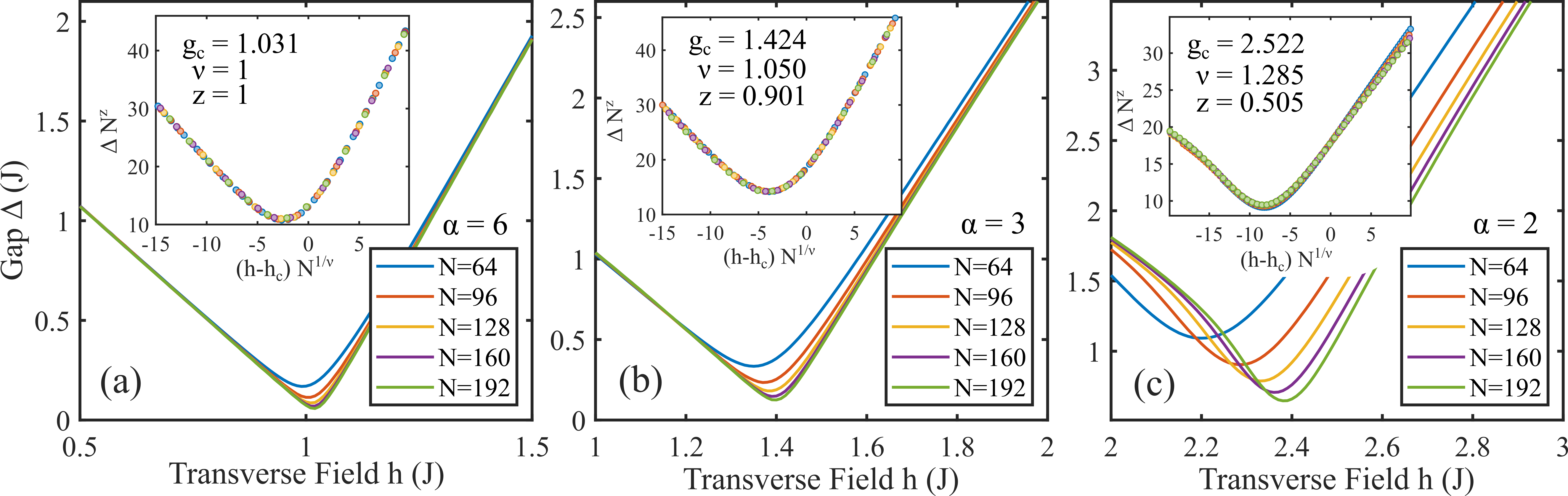}
    \caption{\label{fig:gapCollapse} The energy gap between the ground state and first excited state is shown for spin chains of different length $N$. The insets show the collapsed energy obtained with the indicated critical field $g_c$, critical exponent $\nu$ and critical dynamical exponent $z$. (a) \textbf{$\alpha=6$}. (b) \textbf{$\alpha=3$}. (c) \textbf{$\alpha=2$}.}
\end{figure*}

\begin{figure}
    \centering
    \includegraphics[width=\linewidth]{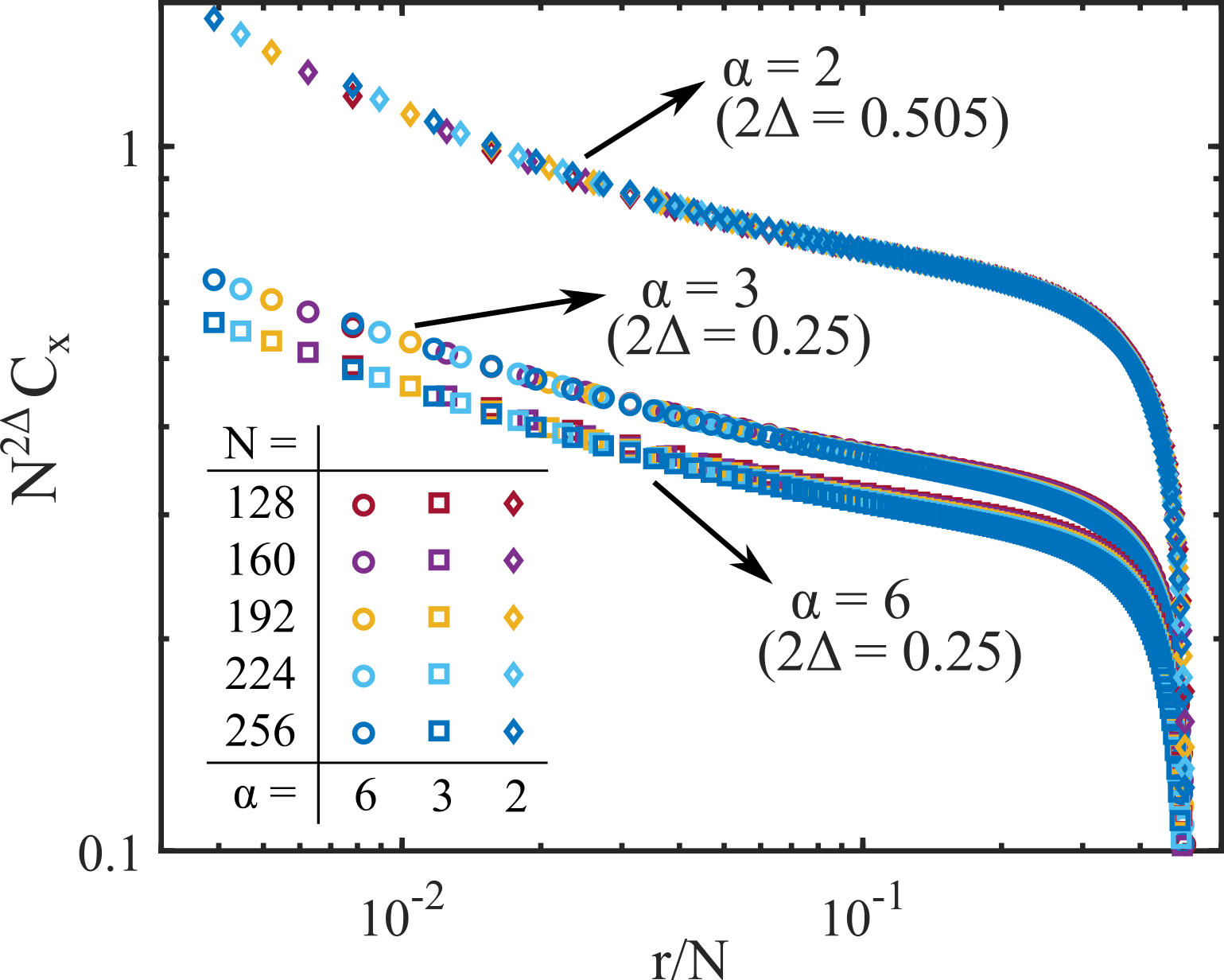}
    \caption{A scaling collapse of the spin correlations calculated with DMRG at criticality for $\alpha = {6, 3, 2}$. The scaling exponents $2\Delta$ are taken from the RG calculations in Ref.~\cite{maghrebi2016}. For $\alpha \geq 3$, $2\Delta = 0.25$ and for $\alpha = 2$, $2\Delta \approx 0.505$.}
    \label{fig:scCollapse}
\end{figure}

\begin{figure}
    \centering
    \includegraphics[width=0.8\linewidth]{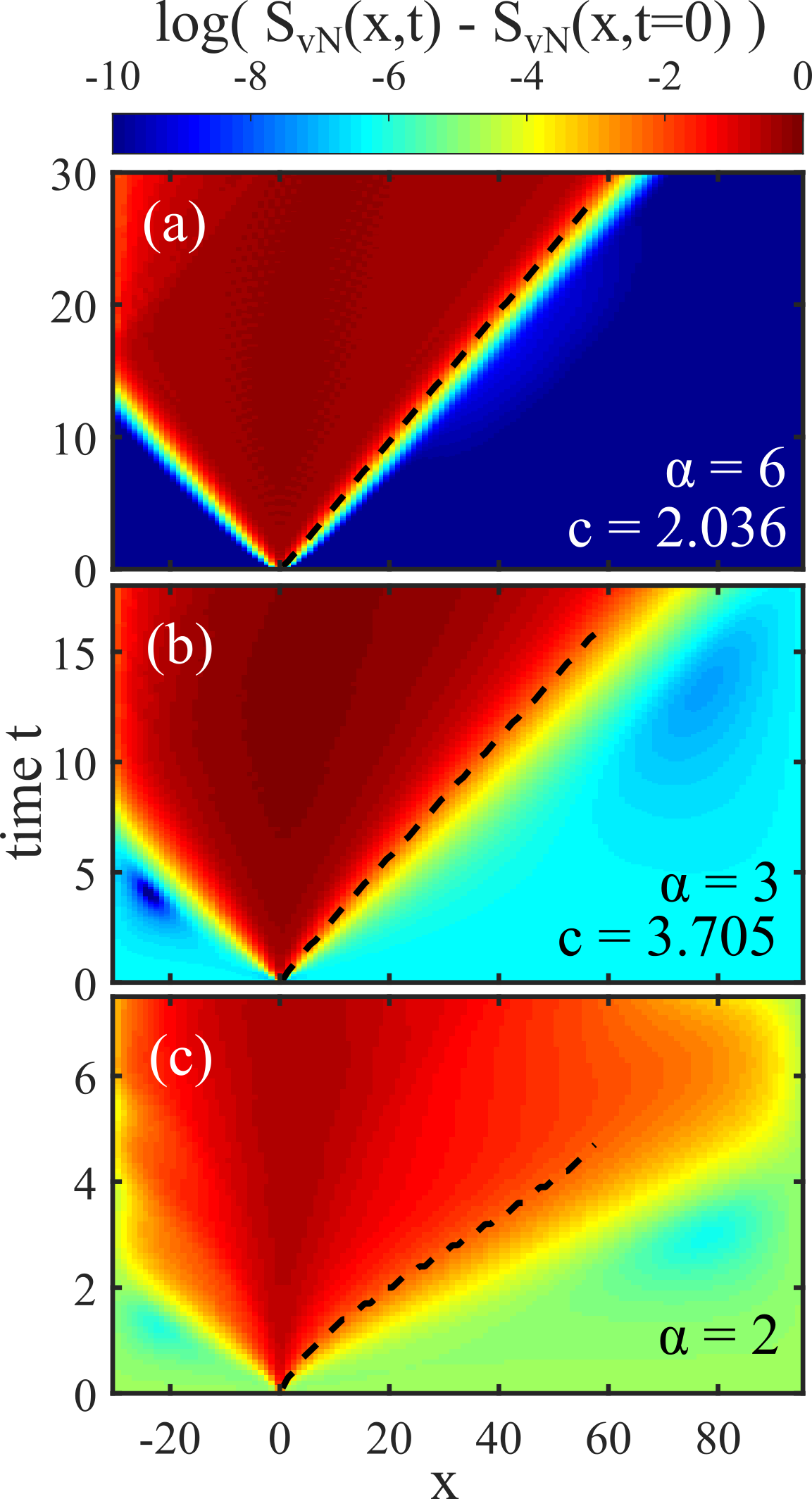}
    \caption{The speed of light $c$ is estimated by doing a local perturbation of the ground state of the LR-TFI model and time evolving with TDVP. The difference $S_\text{vN}(x,t)-S_\text{vN}(x,t=0)$ evolves into a light cone following the perturbation. The black-dotted lines represent the first instance of the difference being larger than a cut-off of $0.1$. (a) In $\alpha=6$, $c\approx2.036$. (b) In $\alpha=3$, $c\approx3.705$. (c) In $\alpha=2$, the light-cone is expected to be logarithmic and no attempt is made to determine the speed of light.}
    \label{fig:sound}
\end{figure}

\section{Properties of LR-TFI Models}
\label{sec:properties}

The critical properties of the LR-TFI models presented in this work are determined by doing a collapse of the energy gap between the ground state and first excited state of the models calculated for different transverse fields $h$ using DMRG. It also serves as a consistency check for the approximation of power-law interacting Hamiltonian by a sum of exponentially decaying Hamiltonians. For $\alpha=6$, we use the dynamical critical exponent $z=1$ based on Ref.~\cite{maghrebi2016}. The best collapse is obtained for the critical field $g_c\approx1.031$ and critical exponent $\nu=1$ as shown in Fig.~\ref{fig:gapCollapse}(b), closely matching previous investigations~\cite{zhu2018,koziol2021}.

For $\alpha=3$, the minimum gap at different $N$ is not consistent with $z=1$ as expected from RG calculations~\cite{maghrebi2016}. A fit of $\min_h \Delta = N^{-z}$ is done to determine $z\approx0.901$. Although this result differs from Ref.~\cite{maghrebi2016}, it is still consistent with the expectation that $\alpha=3$ is at the limit between linear and polynomial Lieb-Robinson light-cones~\cite{kuwahara2020}. The best collapse is obtained for $g_c \approx 1.424$ and $\nu\approx 1.050$ as shown in Fig.~\ref{fig:gapCollapse}(b), also consistent with previous studies \cite{zhu2018,koziol2021}.

For $\alpha=2$, the dynamical critical exponent is chosen to be the corrected mean-field value of $z \approx \sigma/2 + \rho(\sigma)\epsilon^2$ calculated in Ref.~\cite{maghrebi2016}. Here, $\sigma=1$, $\epsilon = 3\sigma/2-1$ and $\rho(\sigma)\approx1/[24(1+\sigma^2)]$, giving $z \approx 0.505$. The best collapse gives $g_c\approx2.522$ and $\nu\approx 1.285$ as shown in Fig.~\ref{fig:gapCollapse}(b), consistent with the previously cited quantum Monte Carlo and DMRG studies \cite{zhu2018,koziol2021}. Out of the three data sets shown in this section, $\alpha=2$ has the worst collapse. We believe this is attributed to the large finite size effects coming from using open boundary conditions with longer range interactions.

The scaling collapse of the spin correlations presented in Sec.~\ref{sec:lorentzCooling} require knowledge of the scaling dimensions of the correlator at criticality. We calculate these critical correlations systems sizes of up to $N=256$ spins, as shown in Fig.~\ref{fig:scCollapse}. We use the system size dependent critical fields $g_c(N)$, which can be estimated by finding $\min_h \Delta$ at $N=128$. The collapse is done using
\begin{equation}
    C_x(r) = N^{-2\Delta} F_C\left(\frac{r}{N}\right),
\end{equation}
where we assume the only relevant length at criticality for these systems is the system size $N$. The collapse works well for the scaling dimensions predicted by the RG calculations in Ref.~\cite{maghrebi2016}. For $\sigma>7/4$, RG predicts that the scaling dimensions of $C_x$ is the same as the free fermion theory. Thus, we use $2\Delta = 0.25$ for $\alpha\geq3$. For $2/3<\sigma<7/4$, the scaling dimension is calculated using an epsilon series expansion giving $2\Delta = 1 - \sigma/2 + \rho(\sigma)\epsilon^2 + O(\epsilon^3)$. In the case of $\alpha = 2$, it gives $2\Delta \approx 0.505$, which collapses the data very well.

The last property that we are interested in is the speed of ``light" $c$ in these systems. The following is inspired by previous work on the light-cone spread of correlations in long-range interacting systems~\cite{hauke2013}. To include the same finite size effects found in the spatiotemporal quenches, we restrict the calculation to chains of $N=128$ spins. The critical ground states of the LR-TFI models with $\alpha={2,3,6}$ are perturbed with the Pauli matrix $\sigma^z$ at site $N/4$ (which we refer to as $x=0$), as shown in Fig.~\ref{fig:sound}. The wavefunction is evolved with the time-dependent variation principle following the perturbation and the block von Neumann entanglement entropy is calculated at each MPS bond at every time step. For $\alpha\geq3$, the difference $S_\text{vN}(x,t)-S_\text{vN}(x,t=0)$ evolves into a linear light cone from which we can estimate the speed of light, as shown in Fig.~\ref{fig:sound}(a-b). The estimate is done by finding when $S_\text{vN}(x,t)$ grows by a certain amount (which we call the cut-off) compared to the ground state. The cut-off is chosen to be $0.1$, a value that closely follows the light cone boundary for $\alpha=3,6$. For $\alpha=2$, the light cone is expected to spread as $t \sim \exp( 3\sqrt{\log{r}})$~\cite{tran2021}, such that no attempt at calculating a speed of light was made.

\section{Formalism for Free Fermions and Solution to Instantaneous Superluminal Quench}
\label{sec:freefermi}

\subsection{Problem Statement}
\label{sec:fermidef}

The problem we would like to solve is specified by the following action and commutation relations:
\begin{widetext}
\begin{align} \label{eq:probdeffermi}
    & S = \int \!dt \int^{L/2}_{-L/2} \!dx \; \left[ \frac{i}{2} \left( \bar{\psi} \gamma^\alpha \partial_\alpha \psi - \partial_\alpha \bar{\psi} \gamma^\alpha \psi \right) - m \Theta(x - v_s t)  \bar{\psi} \psi \right] \nonumber \\
    & \{ \psi_a^{} (x,t), \psi_b^\dagger  (x',t) \} = i \delta(x - x') \delta_{a,b} \;\;\; \text{and}  \;\;\; [\psi_a (x,t), \psi_b (x', t)] = 0. 
\end{align}
\end{widetext}
We set the speed of ``light", $c = 1$. We will work in the Weyl basis wherein $\gamma^0 = \gamma_x$ and $\gamma^1 = -i \sigma_y$, $\sigma_{x,y}$ being Pauli matrices (see section 4.2.4 in Ref.~\cite{fradkin2013field}, and more generally, Ref.~\cite{peskin1995introduction}), $\bar{\psi} = \psi^\dagger \gamma^0$, where $\psi_a (x,t)$ is the field of the Weyl fermion with spinor index $a = 1 , 2$ and $\partial_\mu = ( \partial_t, \partial_x)$. Note that we will use superscript for the spinor index while the subscript will be used to distinguish different solutions of the Dirac equation. The quench occurs locally along a front that propagates towards the right at a fixed, supersonic speed $v_s > 1$. We define the inverse subsonic velocity $u_s \equiv v^{-1}_s < 1$. 

While the Hamiltonian before and after the quench satisfies the usual discrete symmetries associated with free relativistic fermions, imposing a boundary necessitates the breaking of some of these symmetries, see Ref.~\cite{alonso1997diracfermionbox}. We work with `natural' states that forgo parity (P) and charge-conjugate (C) symmetries but keep time-reversal (T) and the combined CPT symmetry. These states also satisfy the condition that the current $c \psi^\dagger \sigma_z \psi$ is zero at the edges of the system. This is affected with the following set of boundary conditions: $\psi_1 (L/2) = \psi_2 (L/2)$, $\psi_1 (-L/2) = -\psi_2 (-L/2)$. One can check for these conditions that $T \psi (x,t) = - \sigma_x \psi^* (x, -t)$ and $CPT \psi (x,t) = - \sigma_z \psi (-x, -t)$ satisfy the same boundary conditions while $P \psi (x,t) = \sigma_x \psi (-x, t)$ and $C \psi (x,t) = \sigma_z  \psi^* (x, t)$ do not.  

\subsection{Method of solution}
\label{sec:fermisolution}

\subsubsection{General principle.}

We work in the Heisenberg picture and describe the field operator prior to the quench $(t < x / v_s)$ by a mode-expansion in terms of the complete set of solutions of the massive Dirac equation, $i \gamma^\mu (\partial_\mu - m) \psi = 0$.  These are positive-frequency particle (or `electron') modes $v_n (x,t) $, and negative-frequency anti-particle (or `hole') modes $\tilde{v}_n = C v_n = \sigma_z v^*_n$, such that
\begin{equation}
    \psi (x, t < x/v_s) = \sum_n \left[ f_n v_n (x,t) + \tilde{f}^\dagger_n \tilde{v}_n (x,t) \right].
\end{equation}

The coefficients $f_n$ and $\tilde{f}_n$ satisfy the usual fermionic anti-commutation relations: all operators anti-commute besides $\{ f_n , f^\dagger_m \} = \delta_{n,m}$ and $\{ \tilde{f}_n , \tilde{f}^\dagger_m \} = \delta_{n,m}$. The initial state is defined via the relation $f_n \ket{0} = 0$ and $\tilde{f}_n \ket{0} = 0$ for all $n$. Note that this amounts to setting the initial state as being the vacuum of hole-like and particle-like excitations, which is the relevant case for a critical system. As before, the above expansion is valid for all times $t < x/v_s$ since this quench occurs on a space-like hypersurface. 

After the quench, the field operator evolves according to the massless KG equation and the mode expansion above is not valid for $t > x / v_s$. To find correlations for subsequent times, we expand the massive modes in terms of the massless modes. We reserve the notation $u_n (x,t)$ and $\tilde{u}_n (x,t)$ for the massless modes, and define 
\begin{align} \label{eq:fermionmatch}
    v_n \big|_{x = v_s t} &= \sum_m \left[ \alpha^*_{n,m} u_m + \beta_{n,m} \tilde{u}_m \right] \big|_{x = v_st}, \nonumber \\
    \tilde{v}_n \big|_{x = v_s t} &= \sum_m \left[ \alpha_{n,m} \tilde{u}_m + \beta^*_{n,m} u_m \right] \big|_{x = v_st}, 
\end{align}
where $\alpha_{n,m}$ and $\beta_{n,m}$ are the Bogoliubov coefficients, and the second equation follows from the first upon application of the charge-conjugate operation. Then, the evolution of the field operator for times $ t > x / v_s$ can be described by the expansion 
\begin{equation}
    \phi ( x, t > x / v_s) = \sum_n \left[ \gamma_n (x,t) f_n + \tilde{\gamma}_n (x,t) \tilde{f}^\dagger_n \right],
\end{equation}
where
\begin{align}
    \gamma_n (x,t) &= \sum_m  \left[ \alpha^*_{n,m} u_m (x, t) + \beta_{n,m} \tilde{u}_m (x,t) \right], \nonumber \\
    \tilde{\gamma}_n (x,t) &= \sum_m \left[ \alpha_{n,m} \tilde{u}_m (x, t) + \beta^*_{n,m} u_m (x,t) \right]. 
\end{align}

\subsubsection{Dirac inner product and normalization of modes.}

We use a coordinate-system invariant normalization scheme for the modes that allows us to determine the Bogoliubov coefficients. We define the Dirac inner product between two solutions $\psi_a $ and $\psi_b$ as
\begin{align} \label{eq:diracinner}
    (\psi_a, \psi_b) = \int d x \sqrt{g} n_\mu J^\mu_{(a,b)} (x), \nonumber \\
    \text{where} \quad J^\mu_{(\psi_a, \psi_b)} = \bar{\psi} \gamma^\mu \phi.
\end{align}
Here, the integral is over all space, $g$ is the determinant of the induced metric on space-like coordinate, $\sqrt{g} dx $ is the covariant volume element, $n^\mu$ is a future-directed time-like unit vector normal to the space-like hypersurface and $J^\mu_{(a, b)}$ is the Dirac current. If $\psi_a$ and $\psi_b$ satisfy the \emph{same} Dirac equation (massive or massless), then it is easy to check that $\partial_\mu J^\mu_{(a,b)} = 0$. Thus, the integral over all space of the charge associated with the current $n_\mu J^\mu$ is constant over time. Note that: 

(a) If the modes $v_n$ and $\tilde{v}_n$ form a complete set of modes according to the Dirac inner product, that is, $(v_n, v_m) = \delta_{n,m}$,  $(v_n, \tilde{v}_m) = 0$, and the mode operators $f_n$ and $\tilde{f}_n$ satisfy the usual fermionic anti-commutation relations, then it can be shown
that the field operators (and its conjugate) satisfy the correct commutation relations as described in Eq.~(\ref{eq:probdeffermi}). 

(b) From its formulation in Eq.~(\ref{eq:diracinner}), it is explicit that the Dirac inner product is invariant under transformation into a coordinate system which admits a separation between time-like and space-like coordinates, that is, the metric is of the form $ds^2 = [N(x,t)]^2 dt^2 - g (x,t) dx^2$. Thus, the normalization relations $(u_n, u_m) = \delta_{n,m}$, $(v_n, v_m) = \delta_{n,m}$, etc. are invariant under such coordinate transformations. 

(c) The above two properties imply that under a Lorentz transformation of the coordinates (without any change in the operators $f_n$, $\tilde{f}_n$), the field operators continue to satisfy the commutation relations in Eq.~(\ref{eq:probdeffermi}) in the \emph{transformed} coordinates, as appropriate for a relativistic field.

(d) The Dirac inner product has the symmetry that $(\psi_a, \psi_b) = (\tilde{\psi}_a, \tilde{\psi}_b)$. Thus, both particle and anti-particle modes follow the same normalization scheme $(u_n, u_m) = (\tilde{u}_n, \tilde{u}_m ) = \delta_{n,m}$. 


\subsubsection{Determination of $\alpha_{n,m}$ and $\beta_{n,m}$}

To determine the coefficients $\alpha_{n,m}$ and $\beta_{n,m}$, we must evaluate the Dirac inner product between modes along the curve $x = v_st$. It is useful to Lorentz-boost into a coordinate frame given by $x' = \gamma_s ( x - u_s t )$, $t' = \gamma_s (t - u_s x)$ with $\gamma_s = 1/\sqrt{1-u^2_s}$, as in this frame, the quench trajectory is simply $t' = 0$. 

The Dirac inner product evaluated at time $t' = 0$ in this frame reads
\begin{equation}\label{eq:Diracboosted}
    (\psi_a , \psi_b) = \int_{-L/2\gamma_s}^{L/2 \gamma_s} dx' \; \psi_a^\dagger \psi_b^{} \big|_{t' = 0}.
\end{equation}

Assuming Eqs.~(\ref{eq:fermionmatch}) hold at $t' = 0$, one may evaluate $(u_n, v_m)$ and $(u_n, \tilde{v}_m)$ to find
\begin{align} \label{eq:alphabetafermi}
    (u_n, v_m)\big|_{t' = 0} &= \sum_m (u_n, \alpha^*_{n,m} u_m + \beta_{n,m} \tilde{u}_m) \big|_{t' = 0} = \alpha^*_{n,m}, \nonumber \\
    (u_n, \tilde{v}_m)\big|_{t'  =0}  &= \sum_m (u_n, \alpha_{n,m} \tilde{u}_m + \beta^*_{n,m} u_m) \big|_{t' = 0} = \beta^*_{n,m}, 
\end{align}
where we used $(u_n, \tilde{u}_m)\big|_{t' = 0} = 0$ and $(u_n, u_m) \big|_{t' = 0} = \delta_{n,m}$. The above suggests that if Eqs.~(\ref{eq:fermionmatch}) are simultaneously satisfiable, then the coefficients $\alpha_{n,m}$ and $\beta_{n,m}$ must be given by Eq.~(\ref{eq:alphabetafermi}). To confirm that these are indeed the correct solutions, we can substitute these solutions into Eqs.~(\ref{eq:fermionmatch}). Using the commutation relations on the field operators (as in Eq.~(\ref{eq:probdeffermi})) at $t' = 0$ directly confirms the validity of the result.

By the methods above, we may also show the inverse expansion at $t' = 0$:
\begin{align} \label{eq:fermionmatchinverse}
    u_m \big|_{x = v_s t} &= \sum_n \left[ \alpha_{n,m} v_n + \beta_{n,m} \tilde{v}_n \right] \big|_{x = v_st}, \nonumber \\
    \tilde{u}_m \big|_{x = v_s t} &= \sum_n \left[ \alpha^*_{n,m} \tilde{v}_n + \beta^*_{n,m} v_n \right] \big|_{x = v_st}.
\end{align}
Using these, one can easily prove that these fermionic Bogoliubov coefficients have the following useful property: 
\begin{align} \label{eq:bogorelationsfermi}
    (u_n, u_m) &= \sum_{a} \left[ \alpha^*_{a,n} \alpha_{a,m} + \beta^*_{a,n} \beta_{a,m} \right] = \delta_{n,m}. 
\end{align}

\subsection{Solution of problem}

\subsubsection{Normalized modes.}

We now provide details of the solution of the problem defined in Eq.~(\ref{eq:probdeffermi}). The massive particle modes are defined as
\begin{align}
    v_{\pm, k} &= \frac{1}{\sqrt{2L}} \left( v_k \pm i v_{-k} \right) \; \text{for} \;\;  k>0, \nonumber \\
    \text{where} \quad v_{k} &= \begin{pmatrix} \cos\! \left( \theta_{k}/2 \right) \\ \sin\! \left( \theta_{k}/2 \right) \end{pmatrix} e^{- i k x + i \Omega_k t}, \nonumber \\
    \tilde{v}_{-k} &= \begin{pmatrix} \sin\! \left( \theta_{k}/2 \right) \\ \cos\! \left( \theta_{k}/2 \right) \end{pmatrix} e^{ i k x + i \Omega_k t},
\end{align}
while the anti-particle modes are defined as
\begin{align}
    \tilde{v}_{\pm, k} &= \frac{1}{\sqrt{2L}} \left( \tilde{v}_k \pm i \tilde{v}_{-k} \right) \; \text{for} \; \;  k>0, \nonumber\\
    \text{where} \quad \tilde{v}_{k} &= \begin{pmatrix} \cos\! \left( \theta_{k}/2 \right) \\ - \sin\! \left( \theta_{k}/2 \right) \end{pmatrix} e^{i k x - i \Omega_k t}, \nonumber \\
    v_{-k} &= \begin{pmatrix} \sin\! \left( \theta_{k}/2 \right) \\ - \cos\! \left( \theta_{k}/2 \right) \end{pmatrix} e^{- i k x - i \Omega_k t}. 
\end{align}
In the above, $\cos (\theta_k / 2) = \sqrt{\frac{1}{2} + \frac{1}{2} \frac{k}{\Omega_k} }$,  $\sin (\theta_k / 2) = \sqrt{\frac{1}{2} - \frac{1}{2} \frac{k}{\Omega_k} }$, and $\Omega_k = \sqrt{m^2 + k^2}$. The modes satisfy the CPT symmetry conserving boundary conditions for $k L = n \pi + \pi /2$, with $n \in [0, 2, 4, ...)$ for modes $v_{+, k}$ and $n \in [1, 3, ...)$ for modes $v_{-, k}$.

An analogous set of massless modes $u_{\pm, k > 0}$ and $\tilde{u}_{\pm, k > 0}$ can be found by setting the mass to zero in the corresponding formul\ae\, for the massive modes. This is enforced by the substitutions $\Omega_k \rightarrow \omega_k = \abs{k}$, $\cos ( \theta_k / 2) \rightarrow 1$, $\sin (\theta_k / 2) \rightarrow 0$. 

It is also useful to note the form of these modes in the Lorentz-boosted frame. The coordinates and momenta are boosted in the usual way, with $k' x - \Omega_{k'} t \rightarrow k'_R x' - \Omega_{k'_R} t'$ and $- k' x - \Omega_{k'} t \rightarrow - k'_L x' - \Omega_{k'_L} t'$. The spinor part is transformed by multiplication with the matrix $\Lambda \equiv -i e^{ \frac{\omega}{2} \cdot \frac{i}{4} [ \gamma^0, \gamma^1 ] } = \begin{pmatrix} 1/\sqrt{\eta} & 0 \\ 0 & \sqrt{\eta} \end{pmatrix}$. Note that $\omega = \text{tanh}^{-1} ( - u_s )$ is the rapidity associated with the Lorentz boost and $\eta = \sqrt{(1+u_s)/(1-u_s)}$ is the usual relativistic Doppler factor. 

\subsubsection{Bogoliubov coefficients.}

The Bogoliubov coefficients can be evaluated by expressing these modes in the Lorentz boosted coordinates and evaluating the Dirac inner product at time $t' = 0$ as per Eq.~(\ref{eq:Diracboosted}) and Eqs.~(\ref{eq:alphabetafermi}). The coefficients read
\begin{widetext}
\begin{align} \label{eq:bogofermi}
    \beta^{\epsilon, \epsilon'}_{k, k'} = (u_{\epsilon, k}, \tilde{v}_{\epsilon',  k'})^* = \frac{1}{2 L} &\bigg[ \frac{\cos\!\left(\theta_{k'}/2 \right)}{\eta} F\!\left( k_R + k'_R \right) - i \epsilon' \frac{\sin\!\left(\theta_{k'}/2 \right)}{\eta} F\!\left( k_R - k'_L \right) \nonumber \\
    &- i \epsilon \; \eta \; \sin\! \left( \theta_{k'} / 2 \right) F\!\left( k'_R - k_L \right) - \epsilon \epsilon'  \; \eta \; \cos\! \left( \theta_{k'} / 2 \right) F\!\left( -k_L -k'_L \right) \bigg] \nonumber \\
    \alpha^{\epsilon, \epsilon'}_{k, k'} = (u_{\epsilon, k}, v_{\epsilon',  k'})^* =  \frac{1}{2 L} &\bigg[  \frac{\cos\!\left(\theta_{k'}/2 \right)}{\eta} F\!\left( k_R - k'_R \right) - i \epsilon'   \frac{\sin\!\left(\theta_{k'}/2 \right)}{\eta} F\!\left( k_R + k'_L \right) \nonumber \\
    &+ i \epsilon \; \eta \; \sin\!\left(\theta_{k'}/2 \right) F\!\left( - k'_R - k_L \right) + \epsilon \epsilon'  \; \eta \; \cos\!\left(\theta_{k'}/2 \right) F\!\left( -k_L + k'_L \right) \bigg]
\end{align}
\end{widetext}
where $F(x) = \frac{L}{\gamma_s}\text{sinc}\!\left(\frac{xL}{2\gamma_s}\right)$ and the Doppler shifted momenta are given by
\begin{align} \label{eq:Dopplermomenta}
    k_{R} &= \gamma_s \left(k - u_s \omega_{k} \right) \in \big[\frac{\pi}{2L\eta}, \infty \big) ,  \nonumber \\
    k_{L} &= \gamma_s \left(k + u_s \omega_{k} \right) \in \big[\frac{\pi\eta}{2L}, \infty \big) , \nonumber \\
    k'_R &= \gamma_s \left(k - u_s \Omega_{k} \right) \in [k'_{0,-}, \infty) , \nonumber \\
    k'_L &= \gamma_s \left(k + u_s \Omega_{k} \right) \in [k'_{0,+}, \infty) ,
\end{align}
with $k'_{0,\pm}=\gamma_s\frac{\pi}{2}(1\pm\sqrt{1+4m^2/\pi^2}$, which correspond to frequencies
\begin{align} \label{eq:Dopplerfrequencies}
    \omega_R &= k_R = \gamma_s \left(\omega_{k} - u_s k \right), \nonumber \\
    \omega_L &=  k_L = \gamma_s \left(\omega_{k} + u_s k \right), \nonumber \\
    \Omega_R &=  \sqrt{k^2_R + m^2 } = \gamma_s \left(\Omega_{k} - u_s k \right), \nonumber \\
    \Omega_L &=  \sqrt{k^2_L + m^2 } = \gamma_s \left(\Omega_{k} + u_s k \right). 
\end{align}

\subsubsection{Infinite-size limit, chiral excitations populations.}

We now work in the infinite-size limit and analyze the creation of excitations from the vacuum. We look at the creation of massless anti-particles from the massive particle modes sitting in the vacuum (and vice versa) since this conversion precisely amounts to the excitation of the system about the vacuum of the massless modes. Noting that the function $F(x) \rightarrow 2 \pi \delta (x)$ in the limit $L \rightarrow \infty$, we find

\begin{align} \label{eq:boundarymatchfermi}
    v_{\epsilon', k'} (t'\!=\!0)&=  \cos\! \left( \frac{\theta_{k'}}{2} \right) \left[ \tilde{u}_{+, k} + \tilde{u}_{-, k} \right]  \bigg|_{k = k^{-1}_R (-k'_R)} \nonumber \\
    &- i \sin\! \left( \frac{\theta_{k'}}{2} \right) \left[ \tilde{u}_{+, k} - \tilde{u}_{-, k} \right] \bigg|_{k = k^{-1}_L (k'_R)} \nonumber \\
    & - i \epsilon' \sin\! \left( \frac{\theta_{k'}}{2} \right) \left[ \tilde{u}_{+, k} + \tilde{u}_{-, k} \right] \bigg|_{k = k^{-1}_R (k'_L)} \nonumber \\
    &+ \text{particle content} \propto u_{\pm, k}. 
\end{align} 

The above is a direct result of the integration over the 4 momentum-conserving delta-functions of the Bogoliubov coefficients and one of these terms, $\propto \delta (-k_L -k'_L)$, does not contribute. Two of these terms are associated with the production of right-movers ($\tilde{u}_{R, k} = \left( \tilde{u}_{+,k}  + \tilde{u}_{-,k}\right)/\sqrt{2}$), and one term is associated with left-movers ($\tilde{u}_{L, k} = -i \left( \tilde{u}_{+,k}  - \tilde{u}_{-,k}\right)/\sqrt{2}$). We focus on the first and second terms since the third term can be shown to be continuously related to the first term, but carries a momentum $k > k_0$ while the first carries momentum $k < k_0$, where $k_0 = k^{-1}_R u_s \gamma_s m$ is of the order of the mass $m$. 

We now evaluate the energy of the system after the quench. Note that the Hamiltonian $H \equiv \psi^\dagger \cdot h \cdot \psi$, with $h \equiv - i \sigma_z \partial_x$. For $\psi (x, t > x / v_s) = \sum_{\epsilon', k'} \left[ \gamma_{\epsilon', k'} (x,t) f_{\epsilon', k'} + \tilde{\gamma}_{\epsilon', k'} (x,t) \tilde{f}^\dagger_{\epsilon', k'} \right]$, and the state being a vacuum of operators $f_{\epsilon', k'}$ and $\tilde{f}_{\epsilon', k'}$, we find 
\begin{widetext}
\begin{align}
    \avg{H} &= \sum \left[ \left( \alpha^{\epsilon_1, \epsilon'}_{k_1, k'} \right)^* \!\left( \alpha^{\epsilon_2, \epsilon'}_{k_2, k'} \right) \!\left( \tilde{u}^\dagger_{\epsilon_1, k_1} \!\cdot h \cdot \tilde{u}_{\epsilon_2, k_2} \right) \right]  + \sum \left[ \left( \beta^{\epsilon_1, \epsilon'}_{k_1, k'} \right) \!\left( \beta^{\epsilon_2, \epsilon'}_{k_2, k'} \right)^* \!\left( u^\dagger_{\epsilon_1, k_1} \!\cdot h \cdot u_{\epsilon_2, k_2} \right) \right] + \sim ( \tilde{u}^\dagger \!\cdot u \text{ and } u^\dagger \!\cdot \tilde{u}) \nonumber \\
    &= \sum \tilde{u}^\dagger_{\epsilon, k} \!\cdot h \cdot \tilde{u}_{\epsilon, k} - \sum \left( \sum \beta^{\epsilon, \epsilon'}_{k, k'} \tilde{u}_{\epsilon, k} \right)^\dagger \!\cdot h \cdot \left( \sum \beta^{\epsilon, \epsilon'}_{k, k'} \tilde{u}_{\epsilon, k} \right) + \sum \left( \sum \beta^{\epsilon, \epsilon'}_{k, k'} u_{\epsilon, k} \right)^\dagger \!\cdot h \cdot \left( \sum \beta^{\epsilon, \epsilon'}_{k, k'} u_{\epsilon, k} \right) \nonumber \\
    &\approx \sum_k [ 1 - N^F_L (k) ] (-k) \tilde{u}^\dagger_{L, k} \tilde{u}_{L,k} + \sum_k [ 1 - N^F_R (k) ] (-k) \tilde{u}^\dagger_{R, k} \tilde{u}_{R,k} + \sum_k N^F_L (k) \; k \; u^\dagger_{L, k} u_{L,k} + \sum_k N^F_R (k) \; k \; u^\dagger_{R, k} u_{R,k}. \nonumber \\
\end{align}
\end{widetext}
In the above, all indices within any brackets are assumed to be summed over. After the first equation, we neglect the oscillating (in time) terms of the form $\tilde{u}^\dagger \cdot u$ and $u^\dagger \cdot \tilde{u}$ that are expected to average out due to the integral over momenta. We also used Eq.~(\ref{eq:bogorelationsfermi}) to eliminate the $\alpha$ coefficients in favor of the $\beta$ coefficients. The last equation follows by substituting the result of Eq.~(\ref{eq:boundarymatchfermi}) into the second equation while neglecting time-dependent terms of the form $\tilde{u}^\dagger_{\epsilon, k} \tilde{u}_{\epsilon', k'}$ and $u^\dagger_{\epsilon, k} u_{\epsilon', k'}$ that come with $k \neq k'$. The chiral populations read
\begin{align} \label{eq:chiralpopfermi}
    N^F_L (k) &= \frac{\Omega_{k_L} - \omega_{k_L}}{2 \Omega_{k_L}}, \nonumber \\
    N^F_R (k) &= \frac{\Omega_{k_R} - \omega_{k_R}}{2 \Omega_{k_R}},
\end{align}
where $k_R = k / \eta$ and $k_L = \eta k$. 

\begin{figure}
    \centering
    \includegraphics[width = \linewidth]{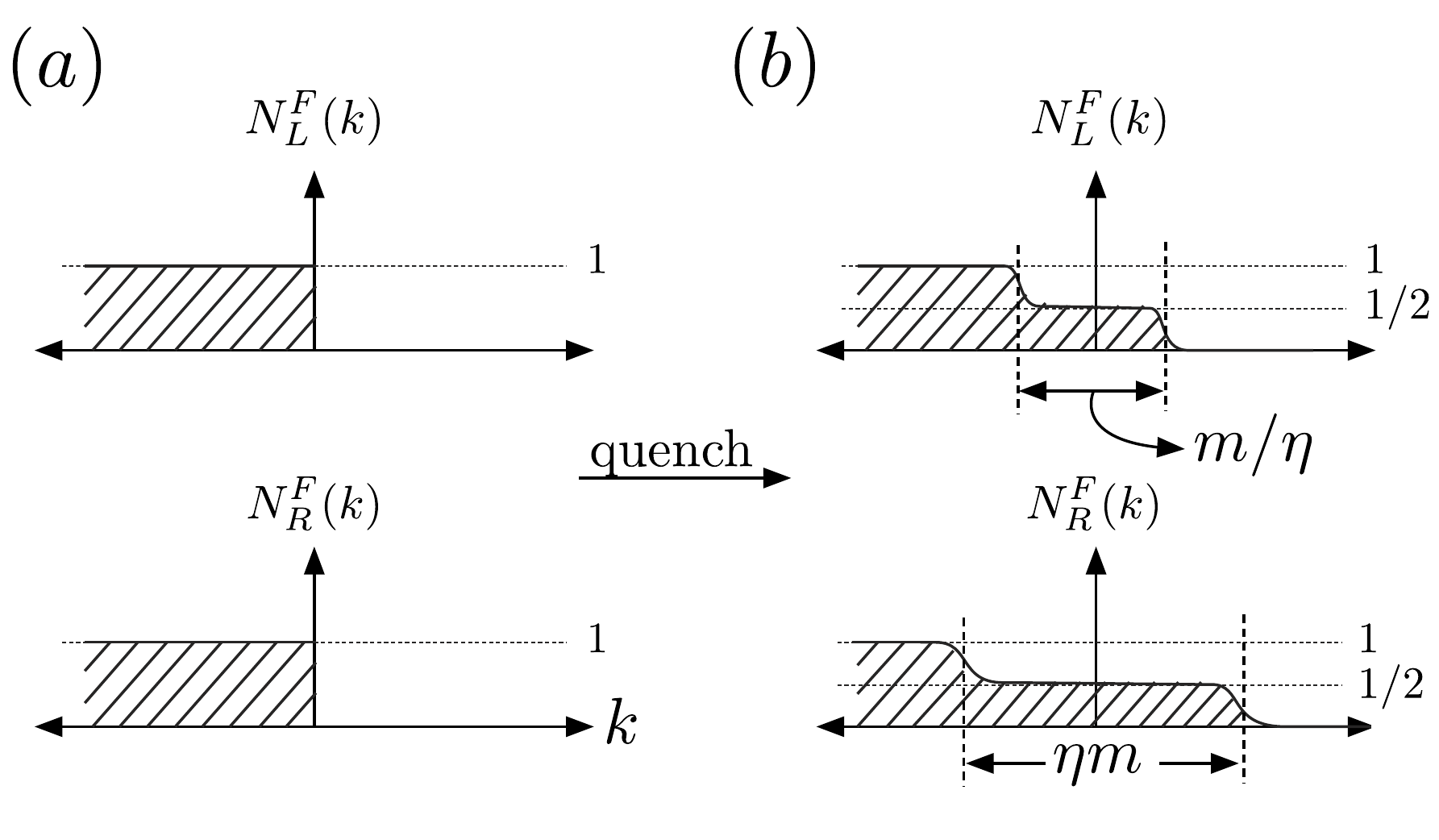}
    \caption{(a) Population of fermions before the quench. The particle states are unoccupied while the hole/anti-particle states are fully occupied. (b) The population of hole and particle states is changed after the quench for momenta $k < m / \eta$ for left-moving modes and $k < \eta m$ for right-moving modes. In the limit $v_s \rightarrow 1^+$, $\eta \rightarrow \infty$, and we see that the left-moving modes are left entirely unexcited.}
    \label{fig:fermipop}
\end{figure}

There is a cooling effect for fermions because the population starts decaying at a Doppler-shifted energy scale: $N^F_L (k) \sim 1/2$ for $k \ll m / \eta$ while $N^F_R (k) \sim 1/2$ for $k \ll \eta m$, and both decrease as $1/k^2$ for larger $k$. The population of fermions before and after the quench is illustrated in Fig.~(\ref{fig:fermipop}). The factor of $1/2$ occurs due to spinor overlap between the massive and massless modes: at $k = 0^+$, the massless modes have a spinor wave-function $( 1, 0 )^T$ or $(0, 1)^T$, but the massive modes have a wave-function $\frac{1}{\sqrt{2}} ( 1, \pm 1)^T $. Thus, the overlap cannot exceed $1/2$ due to Pauli exclusion. Another important difference is that the above result has a UV singularity $\sim 1/k^2$. Thus, the result corresponds to a UV singularity even in $d = 1$. This UV singularity can be eliminated by adding a time-scale to the quench.

\subsection{Quench with a finite time-scale}
\label{sec:fermionfinitetime}

In this section, we discuss how the $1/k^2$ UV singularity is removed by adding a time-scale to the quench. This problem has been analyzed in the context of particle production in inflationary cosmology~\cite{Chungfermionproduction}. There, the metric undergoes a scale change that is equivalent to a scaling of the mass as it is the only term that breaks the conformal invariance of the theory. For the sake of completeness, we note the method used there and quote the result relevant for our purposes.  

We again study the case where the quench occurs uniformly in all space and impose periodic boundary conditions. A generic solution to the time-dependent Dirac equation of motion is then given by 
\begin{align} \label{eq:adiabaticansatzfermi}
    \chi_{k} (t) = \frac{1}{\sqrt{L}} \big[&  \alpha_{k} (t)
    \begin{pmatrix}
        \cos ( \theta_{k} (t) /2 ) \\ \sin ( \theta_k (t) /2 ) 
    \end{pmatrix}
    e^{- i \int^t \omega_{k} (t') dt'}  \nonumber \\
    +& \beta_{k} (t)
    \begin{pmatrix}
        \sin ( \theta_{k} (t) /2 ) \\ \cos ( \theta_k (t) /2 ) 
    \end{pmatrix}
     e^{ \int^t \omega_{k} (t') dt'} \big], 
\end{align}
where $\alpha_{k} (- \infty) = 1$, $\beta_{k} (-\infty) = 0$, $\omega_{k} (t) = \sqrt{m^2(t) + \abs{k}^2}$ and the angle $\theta_k (t)$ is decided by the instantaneous frequency of the mode $\omega_k (t)$. Note that, $\chi_{k} (t \rightarrow -\infty)$ reduces to the massive mode solution $v_k$, while in the $t \rightarrow \infty$ limit it is a linear combination of massless particle and anti-particle solutions with momentum $k$. We are interested in $\abs{\beta_{k} (t = +\infty)}^2$ which is the population of excitations at momentum $k$ after the quench is over.  

Plugging in the ansatz of Eq.~(\ref{eq:adiabaticansatzfermi}) into the time-dependent Dirac equation of motion, we find 
\begin{align}
    \d{\alpha_k}{t} &= - \beta_{k} \frac{k \; dm/dt }{2 \omega^2_k} e^{2 i \int^t \omega_{k}(t') dt'} , \nonumber \\
    \d{\beta_k}{t} &= \alpha_{k} \frac{k \; dm/dt}{2 \omega^2_k} e^{-2 i \int^t \omega_{\vs{k}}(t') dt'}.
\end{align}
We assume $\beta_{k} (t) \ll 1$ and $\alpha_{k} (t) \approx 1$ (justified a posteriori) and solve for $\abs{\beta_{k}}^2$. We now use the result from Ref.~\cite{Chungfermionproduction} for this integral: it is approximated using the steepest descent method and is a reasonable approximation when $\tau^{-1} \lesssim \omega_k (t)$ --- thus it is valid for $\tau^{-1} \lesssim m$ and $k \gtrsim m$. The result for the mass $m (t) = m f(-t/\tau) = \frac{1}{2} + \frac{1}{2} \text{tanh} (-t/\tau)$ is
\begin{equation}\label{eq:timesuppressfermi}
    N_{k} \underset{k \gg \tau^{-1}, m}{\approx} \abs{\beta_{k}}^2 \approx e^{- 2 m \tau - 2 \frac{k^2}{m} \tau}.
\end{equation}
Thus, the excitation of modes with momentum $k \gtrsim \tau^{-1}$ is suppressed exponentially. The case where the quench occurs non-uniformly, via a space- and time-dependent mass $m(x,t) = m f [ (x - v_s)/ (v_s \tau) ]$ cannot be analyzed exactly due to the lack of momentum conservation. While one can boost to a frame in which the quench does occur uniformly, the boundaries in this frame can provide momentum kicks, making an exact analysis difficult. Nevertheless, we expect that the analysis above should remain valid for modes with momenta $\gg 1/L$ that are not particularly sensitive to the boundaries.

In the boosted frame, the quench occurs as $m (x',t') = m f \left[ - \frac{t'}{ \tau \gamma_s} \right]$. Thus, we expect the inverse of the time-scale in the boosted frame, $\tau'^{-1} = \tau^{-1} / \gamma_s$ to become the relevant energy scale above which excitations are suppressed \emph{in the boosted frame}. In the laboratory frame, this implies a population
\begin{align}
    N^F_\theta (k) &\approx \frac{1}{2} \; \; \text{for} \; \; k \ll \frac{m}{\gamma_s \eta (\theta)}
\end{align} 
and decaying exponentially in the opposite limits. To remind the reader, for left-moving modes, $\theta = 0$, $\eta (\theta) = \eta$, and for right-moving modes, $\theta = \pi$, $\eta (\theta) = 1/\eta$. Thus, the energy density of the left- and right-moving modes is given by
\begin{equation}
    \epsilon_\theta (\tau) \bigg|_{\tau^{-1} \approx m} \propto \int^{\frac{m}{\gamma_s \eta (\theta)}} \frac{k}{2} dk \propto \frac{m}{L_m} \frac{1}{( \gamma_s \eta(\theta))^2},
\end{equation}
where $L_m = m/c$ has dimensions of length.

\section{Subluminal Cooling}
\label{sec:sublumCooling}

In spatiotemporal quenches in the LR-TFI models with $z=1$, the energy density shows a minimum when the speed $v$ of the front approaches the speed of light. The superluminal case is studied in appendix~\ref{sec:freefermi}. Here, we offer a classical argument for the existence of this minimum due to relativistic effects in the subluminal case. In the following, we assume that $v<c$. We note here as well that in the limit $v \rightarrow 0$, the energy density will again be suppressed, this time because it corresponds to the adiabatic limit.


The modes excited by a subluminal moving quench front interact with both fronts throughout the entirety of the quench---the excitations released from, say, the right-moving quench front will eventually bounce off the left-moving quench front and interact again with the right-moving front. Classically, the incoming wave will be transmitted through and reflected by the front. The moving front forces moving boundary conditions that change the transmitted and reflected waves frequencies and momenta. The following is based on previous work on front induced wavepacket engineering~\cite{gaafar2019,stepanov1993}. The fields on both sides of the moving front are given by
\begin{align}
    \phi_-(x,t) &= a_\text{in}e^{-i(\omega_\text{in}t-k_\text{in}x)} + a_r e^{-i(\omega_r t+k_r x)} \\
    \phi_+(x,t) &= a_t e^{-i(\omega_t t-k_t x)}
\end{align}
where $a_i$ are the amplitudes, $\omega_i$ the frequencies and $k_i$ the wavevectors of the incoming, reflected, and transmitted waves. The boundary conditions at $x=vt$ are $D\phi_+=D\phi_-$ where $D=\{1,\partial_t,\partial_x\}$ leading to the phase matching conditions
\begin{align}
    \frac{\omega_r}{\omega_\text{in}} &= \frac{1-v/V_\text{in}}{1+v/V_r} \label{eq:refl}\\
    \frac{\omega_t}{\omega_\text{in}} &= \frac{1-v/V_\text{in}}{1-v/V_t} \label{eq:transm}
\end{align}
where $V_i = \omega_i/k_i$ is the phase velocity of the different waves.

The modes in the $x<vt$ region are massless and have dispersion $\omega=ck$. The phase velocity is simply $V = c$ such that Eq.(\ref{eq:refl}) becomes
\begin{equation}\label{eq:subDoppler}
    \frac{\omega_r}{\omega_\text{in}} = \frac{1-\beta}{1+\beta} = \frac{1}{\eta^2},
\end{equation}
where $\beta = v/c$, which we recognize as a Doppler red-shifting of the frequency. Therefore, the energy of reflected modes is red-shifted leading to a strong cooling effect as $v\rightarrow c^-$. We note that the Doppler shift factor employed here is different (with $\beta$ inversed to be precise) from the one employed in the superluminal case, but the net result is the same---cooling is strongest close to $v \rightarrow c$. 

The modes transmitted into massive region $x>vt$ acquire a mass such that their dispersion becomes $\omega = c\sqrt{m^2+k^2}$. The phase velocity is $V_t = c/\sqrt{1-m^2/\omega_t^2}$ and is now frequency dependent. Plugging it into Eq.(\ref{eq:transm}) and solving for $\omega_t$ gives
\begin{equation}\label{eq:subTrans}
    \frac{\omega_t}{\omega_\text{in}} = \frac{1+n'\beta}{1+\beta},
\end{equation}
where
\begin{equation}
    n' = \sqrt{ 1 - \frac{\eta^2m^2}{\omega_\text{in}^2} }
\end{equation}
is used to define a cut-off frequency. Indeed, only modes with $\eta m < \omega_\text{in}$ can be transmitted into the massive region. In the limit $v\rightarrow c^-$, $\eta$ diverges and no modes are transmitted. Therefore, the energy density depends only on the modes excited and reflected by the moving quench front. Having shown that the latter is red-shifted, the energy density is dominated by the modes excited by the subluminal moving quench front, a subject of future investigation.

\bibliography{manuscript}

\end{document}